\documentclass[acmlarge]{acmart}
\AtBeginDocument{%
  }

\setcopyright{cc}
\setcctype{by}
\acmJournal{IMWUT}
\acmYear{2026} \acmVolume{10} \acmNumber{3} \acmArticle{126}
\acmMonth{9} \acmDOI{10.1145/3831646}

\usepackage{amsmath}
\usepackage{xspace}
\usepackage{subcaption}

\usepackage{multirow} %tabular cells spanning multiple rows
\usepackage{color} %colour management
\usepackage{enumitem} 

\usepackage{xspace}
\usepackage{xcolor, colortbl}% http://ctan.org/pkg/xcolor

\newcommand{\m}{\textit{M=}}

\newcommand{\sd}{\textit{SD=}}

\newcommand{\F}[3]{$F({#1},{#2})={#3}$}
\newcommand{\p}{\textit{p=}}

\newcommand{\pminor}{\textit{p$<$}}

\begin{document}

\title{\textsc{BlurDriving}: Investigating How Personalized Blur Techniques Impact Drivers’ Performance in Virtual Reality}

\author{Yuan Li}
\orcid{0009-0005-8013-0896}
\affiliation{%
  \institution{The University of Tokyo}
  \state{Tokyo}
  \country{Japan}
}
\email{liyuan@keio.jp}

\author{Mark Colley}
\orcid{0000-0001-5207-5029}
\affiliation{%  
  \institution{UCL Interaction Centre}
  \city{London}
  \country{United Kingdom}
}
\email{m.colley@ucl.ac.uk}

\author{Xinyue Gui}
\orcid{0000-0001-6541-224X}
\affiliation{%  
  \institution{The University of Tokyo}
  \city{Tokyo}
  \country{Japan}
}
\email{xinyueguikwei@gmail.com}

\author{Cristian Rendon-Cardona}
\orcid{0000-0003-1306-6255}
\affiliation{%
  \institution{Team ARAI, Université Paris-Saclay, CNRS, LISN}
  \state{Paris}
  \country{France}
}
\email{cristian camilo.rendon-cardona@universite-paris-saclay.fr}

\author{Pascal Jansen}
\email{pascal.jansen@uni-ulm.de}
\orcid{0000-0002-9335-5462}
\affiliation{%
  \institution{Institute of Media Informatics, Ulm University}
  \city{Ulm}
  \country{Germany}
}

\author{Christian Sandor}
\orcid{0000-0002-3990-2728}
\affiliation{%
  \institution{Team ARAI, Université Paris-Saclay, CNRS, LISN}
  \state{Paris}
  \country{France}
}
\email{cristian.sandor@universite-paris-saclay.fr}

\author{Takeo Igarashi}
\orcid{0000-0002-5495-6441}
\affiliation{%
  \institution{The University of Tokyo}
  \city{Tokyo}
  \country{Japan}}
\email{takeo@acm.org}

\renewcommand{\shortauthors}{Li et al.}

%\newcommand{\Gui}[1]{{\color{red}#1}}

\begin{abstract}
Distracted driving remains a major safety concern, motivating approaches that aim to reduce visual overload before attention breaks down. However, visual overload varies across individuals, making it difficult to determine appropriate interventions for each driver. We investigate whether controllable visual blur can simplify the driving scene and mitigate distraction. To address this challenge, we propose \textsc{BlurDriving}, a target-selective, distance-aware blur system in a Virtual Reality (VR) urban driving simulator, and employ a Human-in-the-Loop Multi-Objective Bayesian Optimization (HITL-MOBO) framework to personalize blur configurations. Across two VR user studies, we evaluated driving under normal conditions in Study 1 and under cognitively demanding conditions in Study 2. We found that personalization revealed strong individual differences in blur preference but did not lead to significant improvements in objective driving performance compared to a no-blur baseline. Qualitative feedback revealed polarized responses: some drivers reported improved focus, while others experienced uncertainty, fatigue, or discomfort. These findings suggest that visual blur is not universally effective. Instead, its benefits depend on individual perceptual strategies and tolerance for visual uncertainty. This work highlights the limits of personalized visual simplification in safety-critical driving and informs adaptive in-vehicle interface design.

\end{abstract}




\begin{teaserfigure}
  \includegraphics[width=\textwidth]{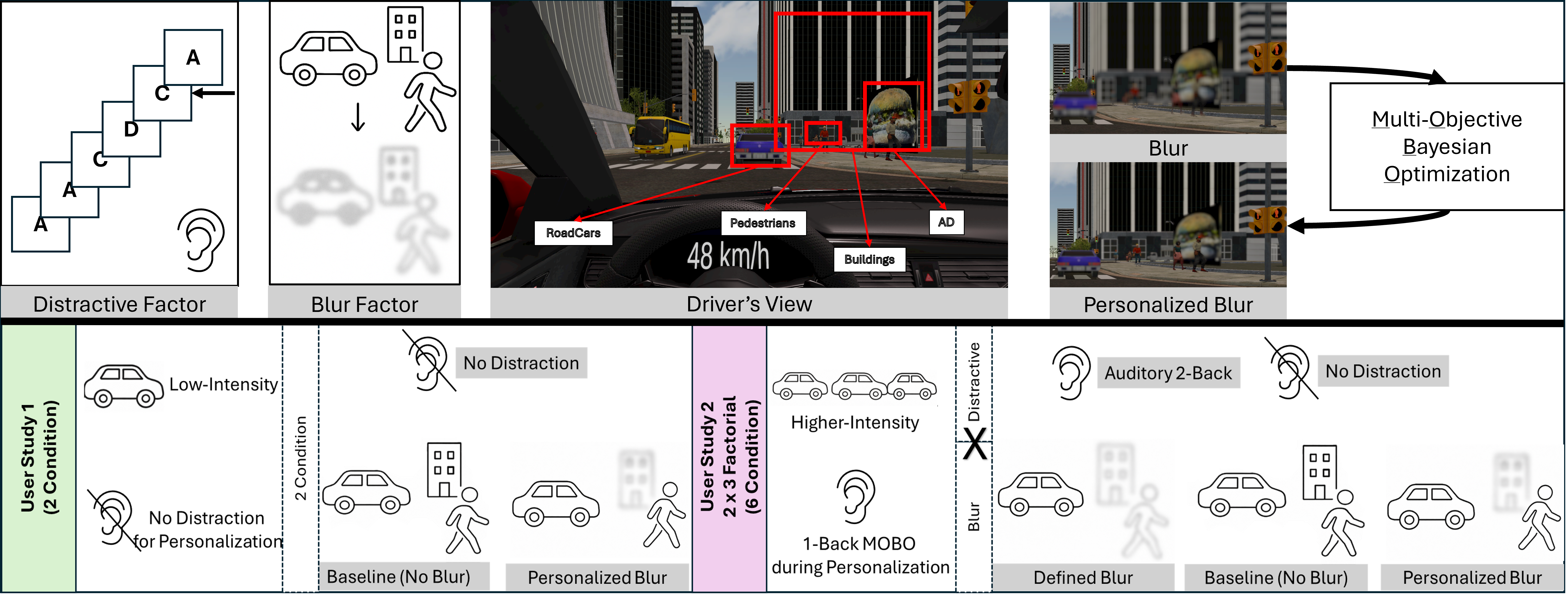}
  \caption{We use human-in-the-loop multi-objective Bayesian optimization (MOBO) to personalize blurring of elements in the driving scene. In the first virtual reality (VR) user study, we explore how personalized blur affects driver performance in a low-traffic, no-distraction environment. In the second VR user study, we explore how blur affects drivers in a cognitively demanding, distracted situation using a 1-back task for the MOBO process and a 2-back task for the evaluation.}
  \Description{}
  \label{fig:teaser}
\end{teaserfigure}

\maketitle

\section{Introduction}
Driving safety remains a major public concern worldwide. According to the World Health Organization (WHO), over 1.19 million people die each year as a result of road traffic crashes, and distracted or inattentive driving is among the key contributing factors~\cite{WHOReport}. The causes of unsafe driving are multifaceted, including drunk driving, fatigue, speeding, and distracted driving~\cite{verschuurModelingSafeUnsafe2008}. Unlike alcohol or speeding offenses, distracted driving often emerges gradually and unintentionally, making it harder to recognize and prevent before it becomes dangerous~\cite{stavrinosImpactDistractedDriving2013}.

Distracted driving occurs when a driver’s attention shifts from the primary task of vehicle control to secondary activities, such as checking a navigation system or operating in-vehicle displays~\cite{Drivermonitorreview}. Rather than causing accidents directly, distraction initiates a gradual process: reduced attention leads to delayed reactions or incorrect actions, which then result in loss of control or collisions~\cite{RECASSimulation}.

To avoid collisions, most prior research has focused on mitigating the consequences after distraction has already occurred~\cite{ADAS, FCW, AEB, DMS, IMWUT-internaldesignspace, IMWUT-TurnsMaps}. These include advanced driver assistance systems such as forward collision warning ~\cite{FCW}, automatic emergency braking ~\cite{AEB}, and lane departure warning, as well as driver monitoring systems that track eye gaze, eyelid closure, and head orientation to infer attention levels~\cite{Drivermonitorreview, IMWUT-DrivingMonitor, IMWUT-Insight, IMWUT-safedrive}. Extensive studies examined the effectiveness of such technologies. However, these approaches remain fundamentally reactive. They intervene only after distraction arises, leaving little room to prevent its onset~\cite{stavrinosImpactDistractedDriving2013}. In contrast, our work aims to \textbf{address the root cause} by proactively minimizing the cognitive load before it occurs.

To do so, we explore a method to lower the driver’s cognitive load by blurring objects in the driver's view. Previous studies have shown that distraction is closely associated with elevated cognitive workload~\cite{Driverdistractionreview}. When cognitive demand exceeds a driver’s processing capacity, attentional resources are reallocated away from the driving task, resulting in suboptimal driving performance. To avoid this, one approach is to simplify the visual environment to facilitate information processing~\cite{tsurukawaFilteringVisualInformation2015}. Previous studies on Diminished Reality (DR) and related visual modification techniques have investigated methods such as removing, replacing, or blurring visual elements to mitigate distraction in extended reality environments~\cite{chengUnderstandingDiminishedReality2022, leeDiminishARDiminishingVisual2025}. However, when applied to driving contexts, complete removal or substitution of visual information would impair spatial awareness and compromise safety~\cite{verschuurModelingSafeUnsafe2008}. Therefore, we propose a blurring-based approach, which preserves essential spatial cues while suppressing irrelevant or distracting visual details. However, it is unclear what the effects of blurring are, what the appropriate level of blurring is, and which objects should be blurred. This makes it difficult to find a design that balances visual information and driver performance in a manual design process (e.g., A/B testing), and the balance is likely to vary by individual, as drivers have different abilities, preferences, and needs. 

To address this challenge, we used Human-in-the-Loop (HITL) Multi-Objective Bayesian Optimization (MOBO), an approach to computationally finding trade-offs in a design space in a few user feedback iterations~\cite{jansenOptiCarVisImprovingAutomated2025a, MarkBOExternalCommunication}. HITL-MOBO iteratively identifies Pareto-optimal design parameters by incorporating user feedback at each iteration, progressively refining the design to meet the design objectives (e.g., high driving safety and low cognitive demand). Pareto-optimal means that a design is optimal in the sense that none of its objectives can be improved without worsening at least one other objective~\cite{marler_survey_2004}. HITL-MOBO has already been successfully applied in various domains to solve design optimization problems~\cite{jansenOptiCarVisImprovingAutomated2025a}. %Beyond this, we have also designed a defined blur configuration based on a previous study in driving safety \cite{}. 

We blurred relevant road objects to prioritize important information. By comparing the MOBO personalized blur with other blur approaches, we can further discuss the effectiveness of blur and how to design blur for driving contexts.
This work is guided by the following two research questions (RQs):

\begin{itemize}
    \item \textit{RQ1: How do blur conditions affect driving performance?}
    \item \textit{RQ2: How does cognitive distraction interact with different blur conditions to affect driving performance?}
\end{itemize}

To investigate our RQs, we first developed a Virtual Reality (VR) based driving simulator. We then constructed an urban traffic environment and integrated \textsc{BlurDriving}, a system that enables manipulation of driving-related visual elements of different importance, such as buildings, pedestrians, vehicles, and advertisements. These elements can be adjusted with different blur configurations, including blur radius and intensity. \textsc{BlurDriving} integrates a HITL-MOBO framework that adaptively adjusts each participant’s blur configuration based on behavioral and subjective feedback, learning personalized models to efficiently identify individual blur preferences~\cite{jansenOptiCarVisImprovingAutomated2025a}. 

We conducted two user studies across different driving task loads and traffic intensities. In the first study, we want to explore the effects of blurring techniques under normal driving conditions. We recruited N = 23 participants to investigate how personalized blur affects drivers' performance in a focused driving context. In the second study, we shift our focus to a more complex distracted driving scenario. We used a 3 x 2 (blur $\times$ distraction) factorial design to investigate how blur affects drivers' performance in a high-intensity scenario, with N = 24 participants. We added a secondary cognitive task (auditory n-back) and more intensive road traffic (e.g., more cars) to simulate distracted-driving scenarios. 
For both user studies, no significant difference was observed between the \textit{blur} and \textit{no blur (baseline)} conditions in overall driving performance or subjective experiences, including perceived safety and cognitive load. 
Open feedback results indicated that the effectiveness of visual blur depends strongly on individual preference. 
Future work should therefore focus on individual differences to further investigate how visual simplification techniques influence driving behavior and perceptual comfort.

\medskip

\noindent\textit{Contribution Statement~\cite{Wobbrock.2016}:}

\begin{itemize} 
\item \textbf{Artifact Contribution: Implementation of \textsc{BlurDriving}.} Design and implementation of a blur system to iteratively optimize the blur configuration through HITL-MOBO, enabling the identification of Pareto-optimal design parameters based on user feedback across multiple design objectives. 

\item \textbf{Empirical study that tells us about how the blur affects the driver's performance.} We conducted two within-subjects user studies involving N = 23 and N = 24 participants. To comprehensively investigate our RQs, we examined the effects of varying traffic densities and experimental conditions, including distracted driving and different blur configurations, on performance.
\end{itemize}

\section{Related work}

\subsection{Visual Simplification for Cognitive Load Reduction}
%TOBEUPDATE, this section needs to answer the following question:
%\begin{itemize}
%    \item Why do we use blur as a visual simplification? Why don't we use other DR solutions?
%\end{itemize}

Visual simplification has been widely utilized to mitigate cognitive load in complex environments~\cite{ryuGazecontingentTrainingEnhances2016}. In the context of driving, this simplification must balance reducing distraction with preserving situation awareness. Existing techniques generally fall into three categories: DR, Visual Attenuation, and Visual Degradation (Blur).

DR involves the complete removal or replacement of real-world objects~\cite{moriSurveyDiminishedReality2017a, jansen2026mirage, 10.1145/3543174.3545255}. While effective at eliminating visual clutter, this approach is often too intrusive for safety-critical tasks like driving. \citet{chengUnderstandingDiminishedReality2022} demonstrated that users prefer to maintain a level of awareness of their surroundings rather than experiencing the "void" created by object removal. In a driving scenario, fully deleting objects (e.g., roadside structures) can sever the driver's connection to the physical context, potentially creating dangerous gaps in spatial awareness.

Visual Attenuation attempts to solve this by reducing the saliency of objects through transparency or desaturation rather than removal~\cite{hongVisualNoiseCancellation2024a}. However, this approach has two flaws for driving. First, transparency or attenuation does not remove high-frequency spatial information (sharp edges and contours). These high-frequency details trigger the ventral visual stream ('vision-for-perception'), compelling the brain to attempt to identify and focus on the object despite its reduced opacity \cite{TobiasDesignSpace, TobiasFlicker}. Second, transparency can introduce perceptual ambiguity regarding depth and solidity \cite{lindemannDiminishedRealitySimulation2017a}, forcing the brain to expend more effort to resolve whether an object is a hazard or a graphic.

This work adopts Visual Degradation (blur) as an appropriate middle ground. Unlike DR, blur preserves the presence and motion of objects; unlike attenuation, it scrubs the high-frequency details (text, sharp edges) that trigger the ventral ("vision-for-perception") stream.
Neurophysiological evidence suggests that the dorsal pathway is not reliant on fine visual acuity~\cite{mannResilienceNaturalInterceptive2010a}. \citet{mannResilienceNaturalInterceptive2010a} demonstrated this resilience in a highly demanding interceptive task, finding that performance remained robust even when visual acuity was significantly degraded. They argue that "dorsally-mediated actions should not be affected by low levels of visual blur". By applying this logic to driving, we can selectively blur distractions (advertisements, complex architecture) to prevent them from grabbing ventral attention (reading/identifying), while relying on the dorsal stream's resilience to blur to maintain safety-critical awareness of their location and motion.

\subsection{Techniques for Mitigating Cognitive Load in Driving}
%TOBEUPDATE

A common strategy to prevent driver distraction from the driving scene is to display information directly in the driver’s line of sight using Head-up Displays (HUDs) or AR-based windshield displays (AR WSDs). These systems project navigation, warnings, or driving data onto the windshield so drivers do not need to look down at a dashboard or phone. Studies show that HUDs can shorten reaction time and help maintain focus on the road~\cite{gabbardGlassDriverChallenges2014b}.
However, AR interfaces also introduce new problems. When too much information is shown, the visual scene becomes cluttered, and drivers may be distracted by the very cues meant to help them~\cite{jansenOptiCarVisImprovingAutomated2025a}. Technical issues, such as poor alignment between virtual and real objects, changes in brightness, and difficulties focusing on objects at different depths, can also compromise safety. \citet{gabbardGlassDriverChallenges2014b} emphasized that AR should support attention rather than compete for it, recommending minimal and well-timed cues.

\citet{lagooMitigatingDriversDistraction2019} combined a HUD with gesture recognition to allow drivers to control messages or navigation without touching a screen. Their simulator study showed fewer rear-end collisions and shorter eyes-off-road times compared to a traditional in-car display. Yet participants also reported higher workload when the interface was visually cluttered, indicating a trade-off between information and distraction.

\citet{ferreiraAugmentedRealityDriving2013a} explored a cooperative “see-through” AR system that shared a video feed between vehicles. By showing what was in front of a truck or bus, the driver behind could see through it and overtake more safely. Although promising, this setup faced practical challenges, including poor image alignment, latency, and added visual complexity that sometimes confused drivers rather than helping them.

To reduce visual overload caused by AR, some researchers propose visual simplification techniques. Instead of adding information, these methods remove or weaken parts of the scene to guide attention.
\citet{eyraudAllocationVisualAttention2015} employed a simulated AR interface to investigate how altering cue saliency affects attention. When general cues were made brighter or larger, drivers looked at them more often but paid less attention to the road itself. In contrast, subtle and task-specific cues helped drivers focus on relevant hazards. \citet{lindemannDiminishedRealitySimulation2017a} went further with DR, experimenting with transparent cockpit views that hide non-essential elements. Their results showed potential for reducing clutter, but also warned that removing too much visual information can harm spatial awareness and lead to uncertainty about nearby objects.

Across these studies, AR and visual simplification share a common goal: helping drivers focus on what matters. Yet both approaches have limits. AR risks adding distraction, while diminished reality risks removing too much context.
To our knowledge, no prior work has explored blurring as a middle-ground technique: one that softens irrelevant areas without hiding them entirely.

% Current method in visual simplification and why the blur technique.

\subsection{Multi-Objective Bayesian Optimization for User Interface Designs}

Blurring in driving scenes is not a single design choice but a multi-parameter UI design problem. Decisions about which objects to blur, blur strength, and distance thresholds jointly affect driving safety and cognitive load. Prior work on driving and Automated Vehicle interfaces shows that visual information density and saliency have strong, but nonlinear, effects on performance and safety, where small parameter changes can shift an interface from supportive to harmful (e.g., over-filtering relevant hazards)~\cite{wickens2008multiple, horrey2007vehicle, colley2023come, kim2009simulated}. As a result, rule-based or fixed blurring strategies may not generalize well across traffic scenes, users, or tasks, and the parameter–outcome relationship is typically unknown at design time.

Several studies in automotive UI contexts, therefore, rely on manual tuning by users or designer-defined presets, for example, adjusting HUD element size, opacity, or color through pilot studies or user preference elicitation~\cite{kim2009simulated}. While such approaches can yield usable "one-size-fits-all" designs, they scale poorly as the number of parameters increases and depend heavily on the designer's subjective intuition. This becomes problematic for perceptual interventions such as blur, where parameter interactions are difficult to reason about and exhaustive exploration is infeasible in user studies. On the other hand, in the user-manually-tuning scenario, users may be overwhelmed by the settings options and unsure how to tune specific values to suit their preferences, as they typically lack design knowledge.

HITL-MOBO addresses these challenges by framing UI personalization as an optimization problem driven by sparse human feedback~\cite{brochu2010tutorial}. Instead of exhaustively testing all configurations, HITL-MOBO iteratively learns a probabilistic surrogate model of the unknown relationship between the design parameters and the design objectives and uses it to propose promising designs that the user evaluates to inform the next design iteration. This motivated its adoption in HCI for preference learning and UI optimization \cite{koyama2020sequential,chiu2020human}. In driving-related contexts, HITL-MOBO has been applied to in-vehicle visualizations on an AR HUD, demonstrating that complex design spaces can be explored with relatively few user evaluations \cite{jansenOptiCarVisImprovingAutomated2025a}.

MOBO is particularly relevant for blurring, as driving-related UI design inherently involves conflicting design objectives, such as minimizing distraction while preserving hazard awareness and perceived safety. MOBO explicitly models these trade-offs and identifies a Pareto front of non-dominated designs rather than collapsing objectives into a single weighted score~\cite{marler_survey_2004}. Alternative optimization methods are less suitable in this setting. Grid search and manual tuning do not scale beyond a small number of parameters, and gradient-based methods require differentiable and noise-free objective functions, which are incompatible with subjective user ratings~\cite{brochu2010tutorial}. Reinforcement learning typically requires substantially more interactions to converge~\cite{chanInvestigatingPositiveNegative2022}, making it impractical and ethically challenging for HITL driving studies. In contrast, MOBO is explicitly designed for low-sample, high-cost evaluations and can provide early, informed configurations during the optimization process. This provides a suitable approach against which predefined or manually tuned blurring strategies can be compared in our study.

\section{Implementation of \textsc{BlurDriving}}
To investigate the effects of blur during driving and to address our RQs, an appropriate blurring mechanism is necessary. Such a mechanism should align with real-world driving safety requirements while allowing for personalization based on user needs.
To this end, we propose \textsc{BlurDriving}, an experimental evaluation system with the following features:

\begin{enumerate}
    \item \textbf{Blur technique:} A Gaussian blurring system that adapts to a maximum safety radius within the driving context and dynamically adjusts blurred objects within the driver’s view.
    \item \textbf{VR-based driving environment:} An urban driving simulation system that provides users with a highly immersive driving experience.
    \item \textbf{Multi-Objective Bayesian Optimization:} An optimization framework that adjusts blur parameters based on users’ subjective and objective feedback.
    \item \textbf{Audio-based secondary task:} For the configuration of user study 2, users are exposed to an auditory n-back task to simulate cognitive distraction.
\end{enumerate}

%In this section, we will introduce the system design and technical details of our systems. The entire experimental system was developed using Unity 6000.2.5f1 as the main development environment.

\subsection{Blur Technique}
\label{BlurTechniqueImpelimentation}
To reproduce a controllable visual blur effect suitable for driving scenes, we implemented a custom Gaussian blur within Unity’s Universal Render Pipeline (URP). Gaussian blur is one of the most widely used blurring algorithms and is known for its computational stability and smooth visual results~\cite{haddadClassFastGaussian1991}. %The design of our blur system met three essential requirements that aligned with the research scope of this study.

%\subsubsection{Target-selective blur by object labels}
The virtual road environment contains multiple potentially distracting elements, such as roadside advertisements, pedestrians, and buildings, organized through layers. 
%To systematically investigate which object categories should be visually simplified, each blur layer in our system specifies a predefined set of Unity \textit{Tags} or \textit{Layers}. 
During rendering, only pixels belonging to designated categories are processed by the blur shader pass. This design enables selective control over the visual emphasis in the driving scene, allowing specific object groups to be dynamically blurred or preserved depending on the experimental condition.

%\subsubsection{Adjustable blur intensity}
Each blur layer \(i\) is assigned an independent blur strength parameter \(\sigma_i\), which defines the spread of the Gaussian kernel applied to the objects within that layer. Let \(I(x, y)\) denote the original image intensity at pixel coordinates \((x,y)\), and \(B_i(x, y)\) denote the blurred output for layer \(i\).
The blur process is defined as a discrete two-dimensional convolution:

\begin{equation}
B_i(x, y) = 
\sum_{u=-k}^{k} \sum_{v=-k}^{k}
I(x+u, y+v)\, G_{\sigma_i}(u, v),
\end{equation}

where \(k\) specifies the kernel radius, and
\(G_{\sigma_i}(u, v)\) is the normalized Gaussian weighting function given by

\begin{equation}
G_{\sigma_i}(u, v) =
\frac{1}{2\pi\sigma_i^2}
\exp\!\left(
-\frac{u^2 + v^2}{2\sigma_i^2}
\right).
\end{equation}

Here, $\sigma_i \in [0,10]$ controls the Gaussian blur strength, with a larger value producing stronger visual smoothing. For optimization, the normalized parameter $u_{\sigma_i}\in[0,1]$ was mapped to the shader parameter as $\sigma_i=10u_{\sigma_i}$.
During the experiment, \(\sigma_i\) can be dynamically adjusted in real time to explore our RQ.

\begin{figure}
    \centering
    \includegraphics[width=1\linewidth]{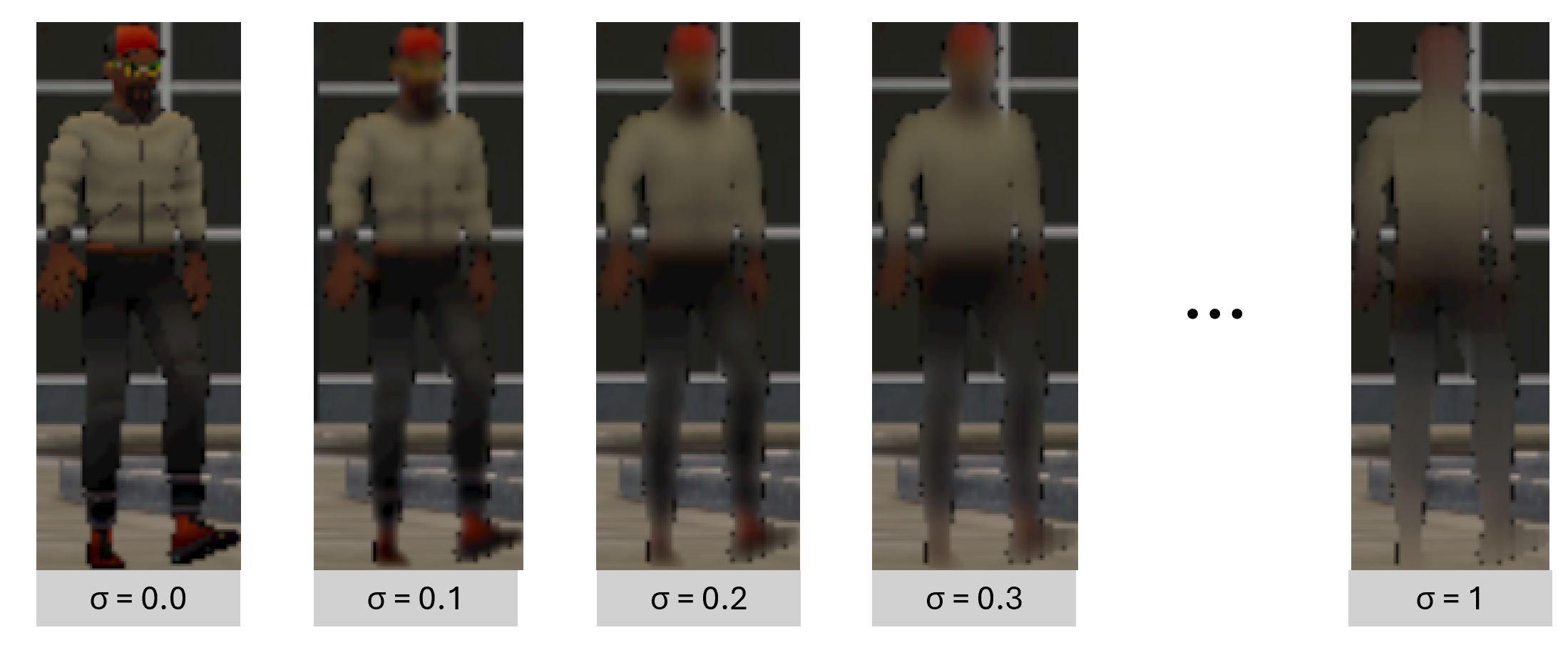}
    \caption{Blur effects for pedestrians at different normalized blur-control values. The displayed labels correspond to the normalized parameter $u_{\sigma}\in[0,1]$, which is linearly mapped to the shader-level Gaussian blur strength as $\sigma=10u_{\sigma}$. The blur strength can be adjusted in real time.}
    \label{fig:placeholder}
\end{figure}

%\subsubsection{Distance-gated blur range}
Finally, the safety distance, i.e., the distance between the ego and other vehicles, is critical when implementing driving simulation systems \cite{ferreiraAugmentedRealityDriving2013a}.
To approximate a safety-related perceptual distance, the blur contribution is modulated by the camera-space distance $d$ between the driver's viewpoint and each object. Two global parameters define the gating range: $R_{\min}$ represents the minimum safety distance within which objects remain sharp, and $R_{\max}$ specifies the cutoff distance beyond which the blur is fully suppressed to preserve global orientation cues. The per-pixel weighting function is given by
\begin{equation}
  w(d) =
    \begin{cases}
      0, & d < R_{min} \\
      1, & R_{min} \le d \le R_{max} \\
      0, & d > R_{max}
    \end{cases}
  \label{eq:blur_weight_band}
\end{equation}
The final blurred intensity for a fragment at distance $d$ is computed as $B'(x,y) = w(d)\,B(x,y) + [1-w(d)]\,I(x,y)$, where $I$ and $B$ denote the original and blurred pixel intensities, respectively. This formulation ensures that blur is selectively applied only to the intermediate distraction zone, while critical elements within the immediate safety zone ($d < R_{\min}$) and distant environmental features ($d > R_{\max}$) remain clearly visible to the driver. A faded effect has been applied between each threshold to reduce cybersickness.

\subsection{Virtual Environment}
We designed and implemented a VR-based simulator using Unity 6000.2.5f1. To simulate realistic driving sensations in the virtual environment, we employed a HORI SPF-004 steering controller~\cite{HORIRacingWheel}. Participants could switch between forward and reverse modes. We used the VIVE Focus Vision, which offers a high-resolution stereoscopic display with a 2448 × 2448-pixel resolution per eye, a 120 Hz refresh rate, and a 120° field of view.
The virtual urban environment for the driving simulation was developed using the Fantastic City Generator\footnote{\url{https://assetstore.unity.com/packages/3d/environments/urban/fantastic-city-generator-157625}; last accessed 31.01.2026} asset.
This asset enables the creation of highly immersive cityscapes and supports procedural generation of urban layouts.

To simulate realistic pedestrian behavior, we employed the Mobile Pedestrian System\footnote{\url{https://assetstore.unity.com/packages/tools/behavior-ai/mobile-pedestrian-system-203706}; last accessed 31.01.2026}.
This system provides configurable walking logic, enabling pedestrians to move along predefined sidewalks and crosswalks in accordance with traffic signals.
%Together, these systems form a dynamic and believable driving environment that closely resembles real-world urban driving conditions.

%\subsection{Virtual Reality Setup}
%The virtual driving experience was implemented using SteamVR version 2.8.0 as the runtime framework within Unity. SteamVR provided a stable and flexible interface between the Unity simulation environment and the VR head-mounted display, enabling precise synchronization of head tracking and rendering.

%For the hardware device, we used the VIVE Focus Vision, which offers a high-resolution stereoscopic display with a 2448 × 2448-pixel resolution per eye, a 120 Hz refresh rate, and a 120° field of view.
%These specifications ensured high visual fidelity and a strong sense of presence during the driving tasks, minimizing potential display latency or motion artifacts.

\subsection{Design of the Multi-Objective Bayesian Optimization Approach}

\begin{figure}[ht]
    \centering
    \includegraphics[width=1\linewidth]{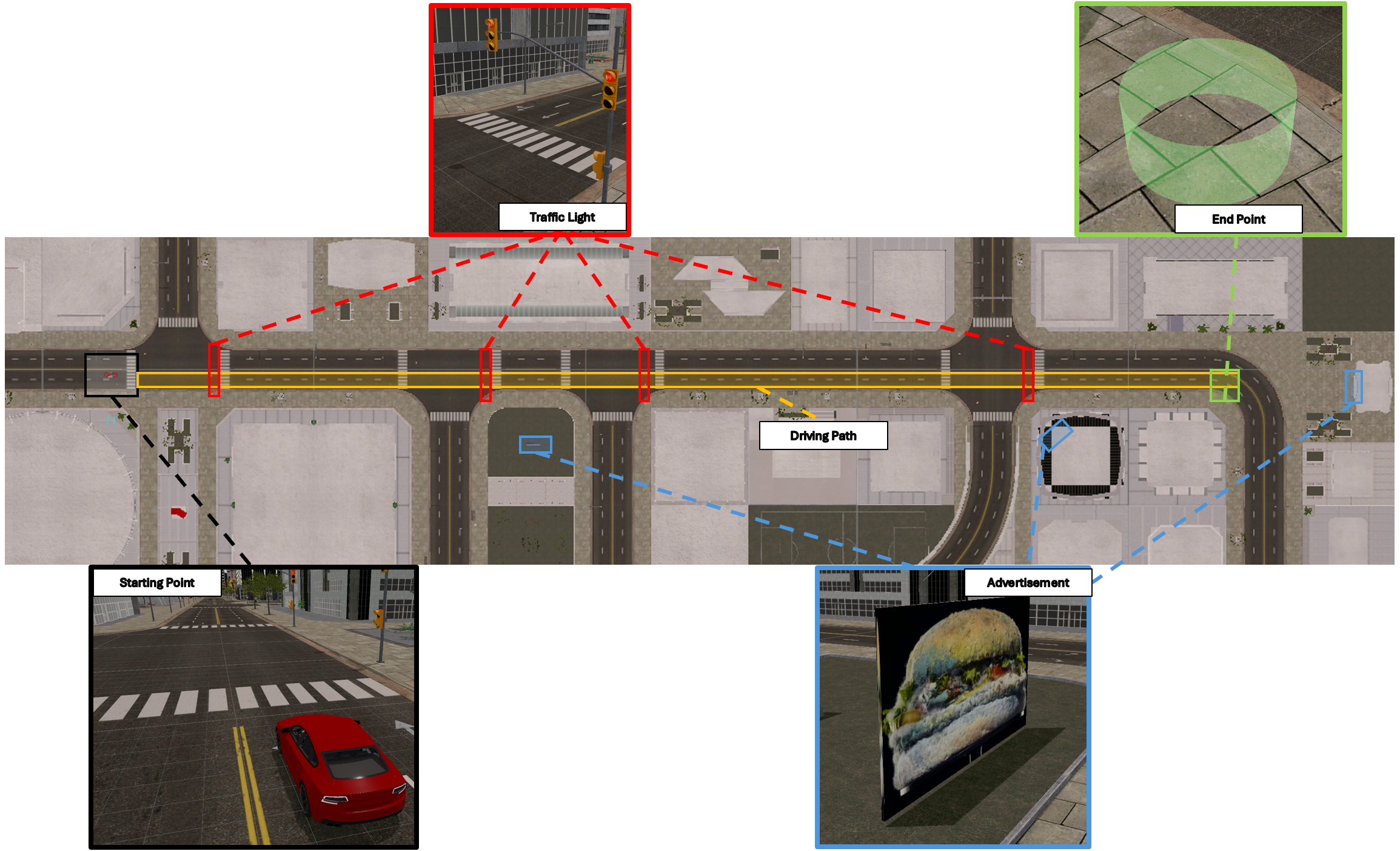}
    \caption{Optimization scenario overview. Participants started at the Starting Point (Left), followed the yellow path, reached the End Point, and then filled in the questionnaire for the next iteration. All participants' objective feedback is being recorded in each iteration.}
    \label{fig:task1}
\end{figure}

We use Bayesian Optimization for the Unity package provided by~\citet{jansen_bayesian_optimization_unity_2025}. We used \texttt{BoTorch}~\cite{balandat2020botorch} version 0.14.3 with a multi-output Gaussian Process using \texttt{qEHVI}.
The optimization process consists of 19 iterations: 13 sampling and 6 optimization. We used a batch size of q = 1 as only one design could be presented during the ride. The HITL process started with 13 sampling iterations using Sobol-sequence samples~\cite{sobol1967} to ensure broad initial coverage of the design space, with identical initial designs shown to all participants. This created a common baseline exploration coverage for the design space. Afterward, we used 6 optimization iterations, during which MOBO adaptively balanced exploration and exploitation to converge toward user-specific designs. 512 Monte Carlo samples were used to approximate the acquisition function. These settings are based on \citet{chanInvestigatingPositiveNegative2022}.

\subsubsection{Design Parameters - Definition of Road Objects}
We chose 6 parameters based on previous studies regarding driving safety. All six optimization parameters were represented on a normalized continuous scale ranging from 0 to 1. For each object-specific blur parameter, the normalized value $u_{\sigma_i}$ was linearly mapped to the shader-level blur strength as $\sigma_i=10u_{\sigma_i}$. Adjusting this parameter allows us to further examine how blurring influences drivers’ attention allocation and safety perception.

\begin{itemize}[leftmargin=1.2em]
    \item \textbf{Global blur radius ($R_{\text{blur}}$):}  
    Defines the overall spatial extent of the Gaussian kernel applied to blurred objects. 
    It controls the maximum blur distance and determines how strongly distant objects are visually simplified. This parameter stands for \(R_{\max}\) in Section \ref{BlurTechniqueImpelimentation}. The default \(R_{\min}\) used during personalization was 30 m. For condition in Study 2, \(R_{\min}\) was set to 40 m., following the safety braking distance in city driving speed limits~\cite{sugiyantoDETERMININGMAXIMUMSPEED2018}.

    \item \textbf{Building blur intensity ($\sigma_{\text{building}}$):}  
    Controls the standard deviation of the Gaussian kernel for architectural elements such as walls, facades, and roadside buildings.

    \item \textbf{Vehicle blur intensity ($\sigma_{\text{vehicle}}$):}  
    Determines the blur strength of non-player vehicles (traffic cars). 

    \item \textbf{Advertisement blur intensity ($\sigma_{\text{ad}}$):}  
    Specifies the blur strength for roadside billboards and commercial signs. 
    These objects are intentionally included to test the effect of reduced visual distraction.

    \item \textbf{Roadside object blur intensity ($\sigma_{\text{roadside}}$):}  
    Defines the blur strength applied to static decorative or environmental objects, 
    such as trees, fences, and benches located near the driving path.

    \item \textbf{Pedestrian blur intensity ($\sigma_{\text{ped}}$):}  
    Controls the blur strength applied to walking pedestrians. 
    
\end{itemize}

\subsubsection{Scenario for Optimization}
Participants drove along a straight road from the left start to a green goal at the far right, passing four signalized intersections along the way. The roadway contained typical urban distractors, including pedestrians, other vehicles, and roadside advertisements. Each traversal from start to goal constituted one iteration. Upon reaching the goal, participants completed a short post-iteration questionnaire that included multiple objective measures, as detailed in the next section.

\subsubsection{Objective Function for Optimization}\label{sec:objectivefunction}

An objective function maps a specific blur configuration to a measurable outcome that the optimizer either maximizes or minimizes. In our optimization, we maximize two objectives: the subjective metric \textit{Perceived Safety}~\cite{faasLongitudinalVideoStudy2020} and the objective metric \textit{Driving Safety Score}. In parallel, we minimize \textit{Cognitive Demand}, which serves as an additional subjective objective.

We employed the following questionnaires to retrieve these metrics after every optimization iteration in the HITL process: We used Raw NASA-TLX Score~\cite{DevelopmentNASATLXTask1988} on a 20-point Likert scale to evaluate the \textit{Cognitive Load} (lower the better) (including six original subscales: Mental Demand, Physical Demand, Temporal Demand, Performance, Effort, Frustration). Participants rated their \textit{Perceived Safety} using four 7-point semantic differentials from 1 (Anxious / Agitated / Unsafe / Timid) to 7 (Relaxed / Calm / Safe / Confident)~\cite{faasLongitudinalVideoStudy2020}.

\textit{Driving Safety Score} is measured following participants' driving performance.
\label{LongitudinalSafetyScore}
The driving safety score is designed to quantify longitudinal driving stability in real time, following established methodologies in telematics-based driving assessment~\cite{NSWYoungDrivers2019, orsiniHighwayDecelerationLane2021}. 
Prior studies have shown that excessive acceleration and braking (often referred to as ``harsh events'') are strongly correlated with increased crash risk and reduced driving safety. 
In our implementation, the score is continuously computed from forward acceleration signals using a sliding window with dwell-time filtering, asymmetric acceleration/braking thresholds, and time-weighted penalty accumulation.

This score integrates both discrete violation detection and continuous safety evaluation during each iteration of the optimization process. Specifically, during each driving trial, three types of rule violations were automatically monitored:
(1) \textbf{Red-light violation} whether the participant crossed an intersection during a red signal;
(2) \textbf{Collision event} whether the vehicle physically collided with another object or pedestrian; and
(3) \textbf{Overspeeding} whether the driving speed exceeded the preset limit of 50~km/h continuously for more than 5 seconds.  
Due to device and latency limitations, transient overspeed events shorter than 5 seconds were not counted as violations. At the end of each iteration, the system determined the objective safety score \(S_{\text{obj}}\) as follows:
\[
S_{\text{obj}} =
\begin{cases}
0, & \text{if any violation (red-light, collision, or overspeed) occurred}, \\[6pt]
S_{\text{LSS}}, & \text{otherwise,}
\end{cases}
\]
where \(S_{\text{LSS}}\) denotes the Longitudinal Safety Score computed from real-time acceleration data (see Section 3).  
In this formulation, the occurrence of any traffic violation immediately nullifies the safety score, whereas compliant driving behavior retains the actual longitudinal smoothness score measured during that iteration. This rule-based objective definition enables the optimizer to balance both binary safety constraints (rule compliance) and continuous behavioral performance (driving smoothness), thereby guiding the search toward safer, more stable blur configurations.

\subsection{Audio-based secondary task} 
We introduce an auditory n-back task as a controlled cognitive factor. The n-back task has been widely used in HCI and psychology, primarily to manipulate participants’ cognitive workload across different experimental settings~\cite{nback->review01, nback->review02, nback->drivingex01, nback->drivingex02, nback->drivingex03, nback->drivingexHazard}. In real-world driving scenarios, drivers are often subject to environmental influences that lead to secondary task engagement, for instance, engaging in conversation with passengers inside the vehicle, which poses potential risks of distraction~\cite{WHOReport}. In the domain of driving safety research, several studies have adopted the n-back task as a cognitive stimulus to induce distracted driving~\cite{nback->drivingex01, nback->drivingex02, nback->drivingex03, nback->drivingexHazard}. The n-back task offers several key advantages~\cite{nback->review01, nback->drivingex02}: (1) it provides a convenient and standardized method to adjust cognitive load intensity; (2) it relies on auditory stimuli and requires only verbal or button-based responses; (3) it minimally interferes with the driver's primary visual-manual driving operations. In our system, participants respond to the auditory n-back task by pressing X (Cross) buttons on the steering wheel, allowing them to complete the task without removing their hands from the driving interface.

\section{User Study 1 - Blur Technique in Low-Intensity Driving Context}
\label{userstudy1}

In User Study 1, we addressed RQ1 by comparing personalized blur with a no-blur baseline in a low-intensity driving context without a secondary distraction task. We employed the HITL-MOBO strategy to optimize blur settings for each participant. We also compared the personalized blur condition against a no-blur baseline. All trials were conducted in a relatively normal driving context, with less crowded traffic and no secondary distraction tasks.
There are two conditions for user study 1:
\begin{itemize}
    \item \textbf{No Blur (Baseline):} Participants driving in the city without blur.
    \item \textbf{Personalized Blur:} Participants driving in the city with a blur profile personalized by HITL-MOBO.
\end{itemize}

All experiments were conducted on a high-end gaming laptop equipped with an AMD Ryzen 9 HX365 CPU, an NVIDIA RTX 5070 Ti GPU, and 32 GB of RAM. All user study was reviewed and approved by the Ethics Review Committee of The University of Tokyo (Approval No. UT-IST-RE-250714).

\subsection{Scenario for User Study}
\subsubsection{Traffic Rules}
The overall traffic system in the virtual driving environment followed a right-hand traffic rule. The scenario is being set up for a downtown city driving scenario with complex visual information. In our experimental scenario, the driving task followed three traffic safety regulations to ensure consistent behavioral evaluation:
(1) participants were instructed not to run red lights, (2) to avoid collisions with pedestrians, vehicles, or any roadside objects, and (3) to maintain their speed below the designated limit of 50 km/h. Most urban areas typically set speed limits within the range of 30–50 km/h~\cite{sugiyantoDETERMININGMAXIMUMSPEED2018, luSimulationbasedPolicyAnalysis2023, elvikHowCanNotion2018}.
Accordingly, we set the speed limit in our driving scenario to 50 km/h to reflect a realistic urban driving environment.
Given that our hardware setup differs from that of a real vehicle, a tolerance margin was introduced for speeding detection. Specifically, a speeding event was only recorded if the participant exceeded the 50 km/h speed limit continuously for more than 5 seconds.
Any traffic violation—including speeding, running a red light, or collisions—was reflected in the participant’s objective safety score, as described in the following section.

\subsubsection{Scenario Design}
\begin{figure}
    \centering
    \includegraphics[width=1\linewidth]{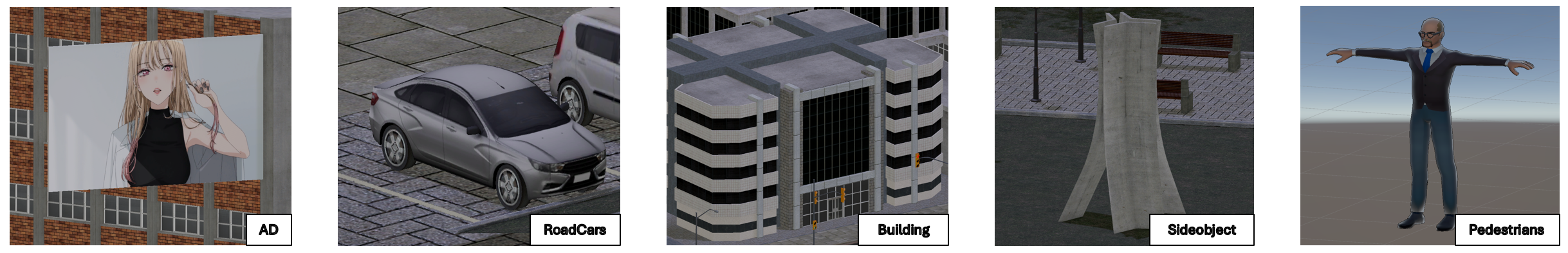}
    \caption{In-scene objects to simulate realistic and complex urban driving scenarios. The models displayed sequentially include advertisements, roadcars, buildings, side objects, and pedestrians.}
    \label{fig:exampleob}
\end{figure}

The driving scenario included pre-generated traffic vehicles operating along a fixed route.
Pedestrians were implemented using the Gley Pedestrian System, which provided realistic walking and crossing behaviors (as also described in Section~3).
To simulate potential sources of driver distraction, we introduced roadside advertisements as visual stimuli.
These advertisements were deliberately designed to be attention-catching and categorized into three thematic types: horror, cute, and food.
Each category was evenly distributed across different experimental environments to ensure balanced exposure among conditions. The surrounding buildings were generated using the Fantastic City Generator to create a realistic urban environment. Additional roadside elements, such as bus stops, sculptures, trees, and utility poles, were included as side objects to enhance environmental richness and spatial realism.
Together, these components were designed to simulate a typical urban driving scene and provide participants with a high sense of presence during the user study. \autoref{fig:exampleob} illustrates several designs of objects included in the driving scene.

\begin{figure}
    \centering
    \includegraphics[width=0.7\linewidth]{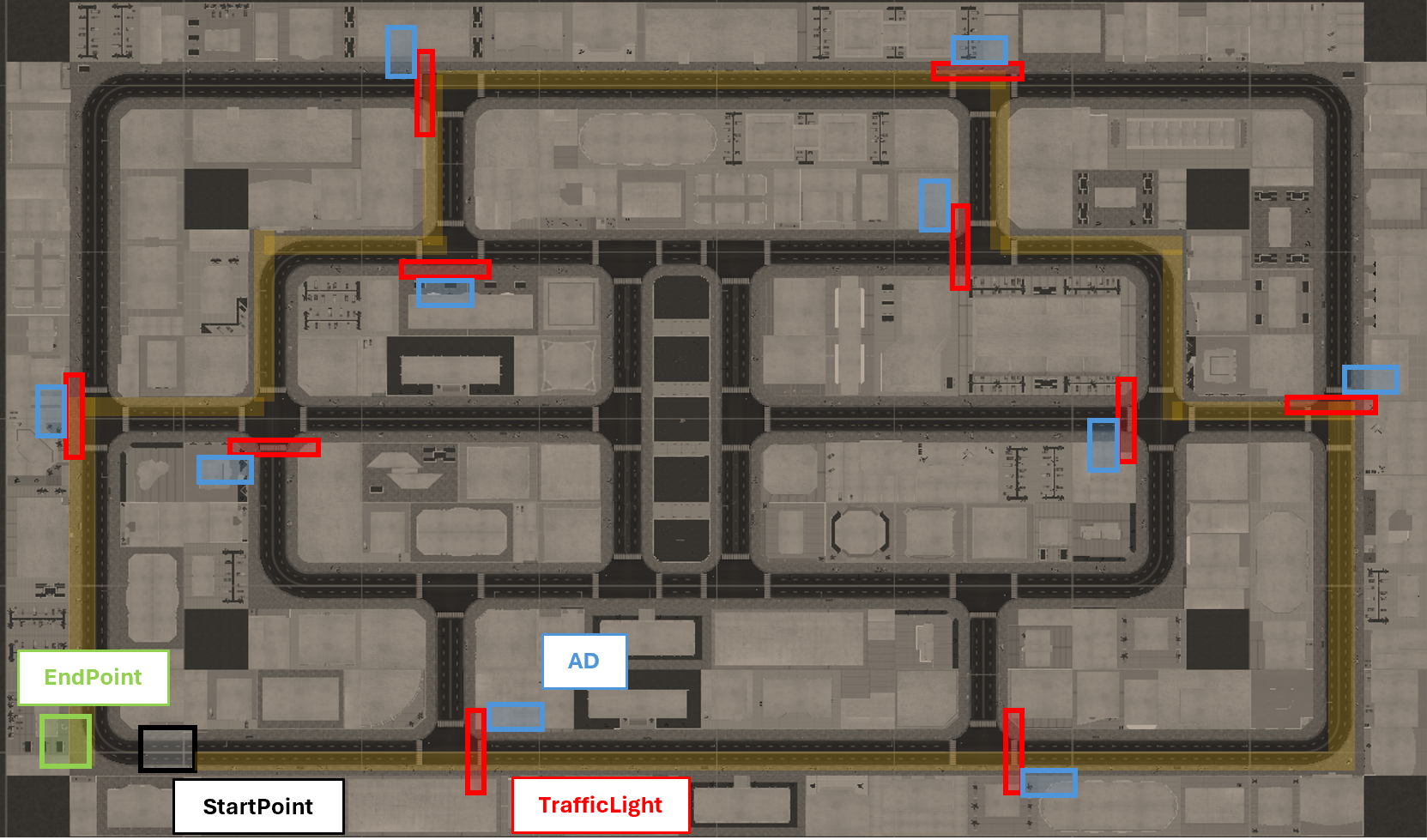}
    \caption{Main Driving Task. After optimization, participants start at the black area in the bottom-left corner and drive along the yellow path until they reach the endpoint (green). Red: traffic light, blue: advertisement or billboard. }
    \label{fig:task2}
\end{figure}

\subsection{Additional Measurements}
\label{study1-measurements}

Beyond the measurements necessary for HITL-MOBO described in Section~\ref{sec:objectivefunction}, including \textit{Perceived Safety}, \textit{NASA-TLX}, \textit{Collisions}, \textit{Red Light Violations}, \textit{Overspeed Duration}, and \textit{Min Safety Score}, we additionally took the following subjective measurements. \textit{Subcales of NASA-TLX \cite{DevelopmentNASATLXTask1988}}, to comprehensively measure participants' cognitive load during the user study. \textit{Blur Perceive Level}, using 7-point semantic differentials from 1 (Not at all) to 7 (Totally) to measure how participants perceived the blur effect during the user study. \textit{Preference Condition}, measuring which condition is most preferred based on their driving experience. \textit{Preference for Safer Driving}, measuring which condition is preferred for safer driving.

%\subsubsection{Subjective Measures}
%\subsubsection{NASA-TLX (Raw TLX Score)}
%he NASA Task Load Index (NASA–TLX) was used to measure the participants’ perceived workload after completing each condition.  
%We adopted the raw TLX score without weighting, which includes six dimensions: mental demand, physical demand, temporal demand, performance, effort, and frustration.  
%Participants rated each dimension on a 20-point scale, and the mean value was computed as the overall workload score.
%\subsubsection{Perceived Safety}
%Participants rated their perceived safety using four 7-point semantic differentials from 1 (anxious/agitated/unsafe/timid) to 7 (relaxed/calm/safe/confident; higher is better).

%\subsubsection{Blur Recognition}
Participants were asked whether they consciously noticed the presence of blur, rated on a 7-point Likert scale from 1 (not perceived at all) to 7 (totally perceived).  
This helped verify whether the applied blur parameters were recognized in our virtual environment.

%\subsubsection{Preference}
At the end of the experiment, participants were asked to indicate their preferred condition (\textit{blur} or \textit{baseline}) and explain the reason for their choice.  
They were also asked which condition they believed contributed more to safe driving.  
These responses were later analyzed to understand individual differences in perception and preference toward visual blur techniques.

After completing both conditions, participants were asked to freely describe their subjective impressions of each driving experience.

%\subsubsection{Objective Measures}

%\subsubsection{Objective Safety Score}
%The \textit{Objective Driving Safety Score} quantified longitudinal driving stability in real time.   It was derived from vehicle acceleration data using the method described in Section \ref{LongitudinalSafetyScore}.  The score ranged from 0 to 100, with higher values indicating smoother, safer driving behavior. The sampling rate is 50Hz.  
%\subsubsection{Collisions}
%Collision events were automatically detected whenever the participant’s vehicle physically contacted pedestrians, other vehicles, or roadside objects.   Each collision was logged as a single event and counted toward the objective safety evaluation.
%\subsubsection{Red Light Violations}
% Red light violations were recorded when the participant crossed an intersection while the traffic signal was red.  
%\subsubsection{Total Overspeed Duration}
% Overspeeding was monitored throughout the driving task.  
% Given that the experimental hardware setup differs from a real vehicle, a 5-second tolerance was applied before registering an overspeed event.  
 %The total overspeed duration (in seconds) was calculated for each trial and used as an indicator of speed control performance.

%\subsubsection{Open Feedback}

\subsection{Participants}
Participants were recruited by the university's social account. 23 (16 male, 7 female, 0 non-binary) participants aged \m{26.98} (\sd{2.82}) took part in our experiment. All participants have a valid driver's license for \m{5.11} (\sd{3.67}) years. Previous VR experiences on a scale from 1 to 5 averaged \m{2.22} (\sd{0.72}).

\begin{figure}[ht]
    \centering
    \includegraphics[width=0.5\linewidth]{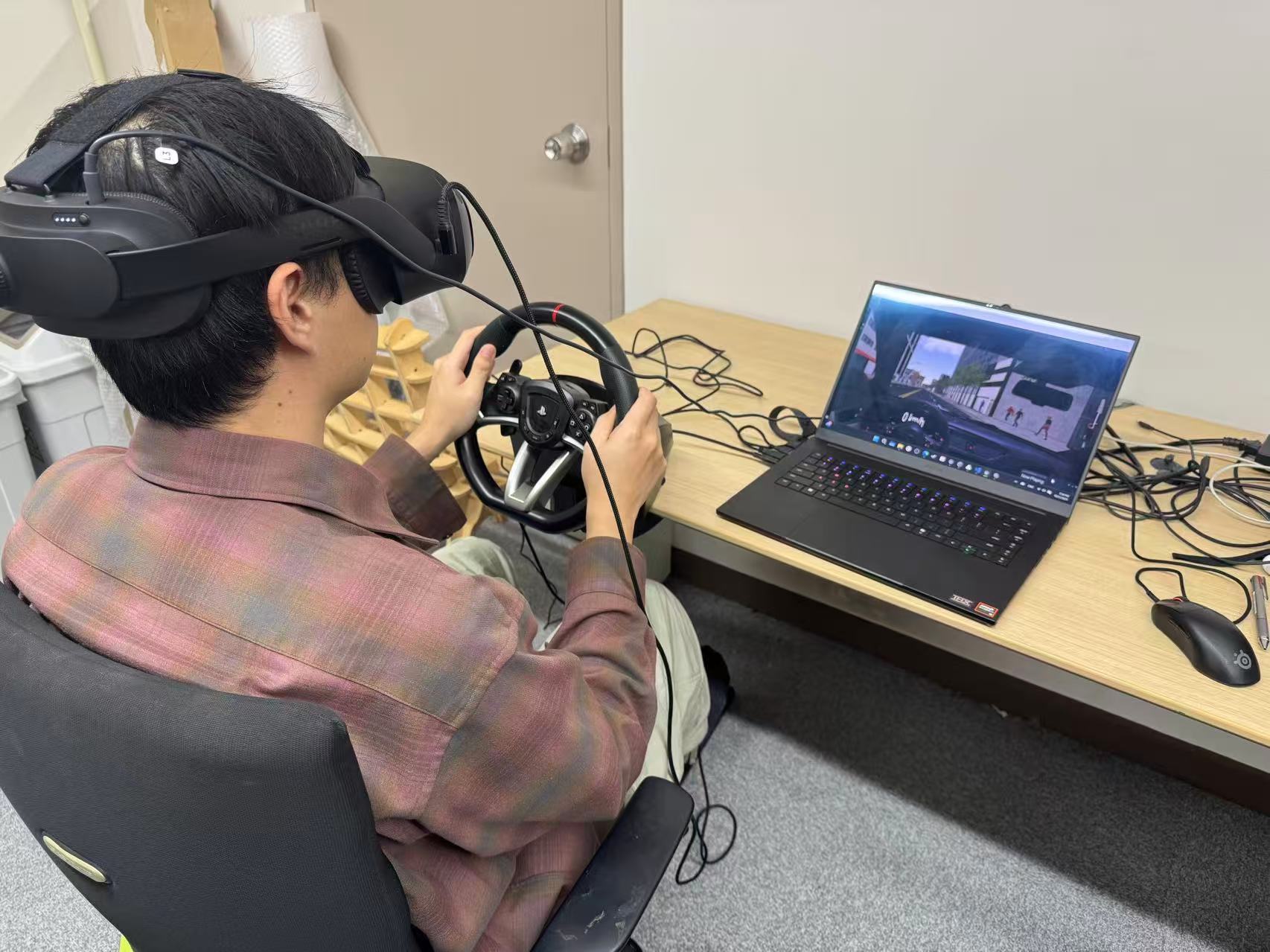}
    \caption{Participants during the user study}
    \label{fig:participant}
\end{figure}

\subsection{Study Procedure}
Participants were recruited and registered for the study through the university’s internal social network. Upon arrival, the instructor explained the study procedure in detail, and participants began the experiment after providing written informed consent. Then, participants practiced 5 minutes (see \autoref{fig:participant}. After practice, participants proceeded to an optimization session, during which a personalized blur configuration was optimized for each individual using HITL-MOBO, while both subjective questionnaire data and objective driving performance metrics were recorded. After optimization was complete, participants proceeded to the main driving task, which consisted of two driving conditions (Personalized Blur and Baseline) in a balanced order. 

\begin{figure}
    \centering
    \includegraphics[width=1\linewidth]{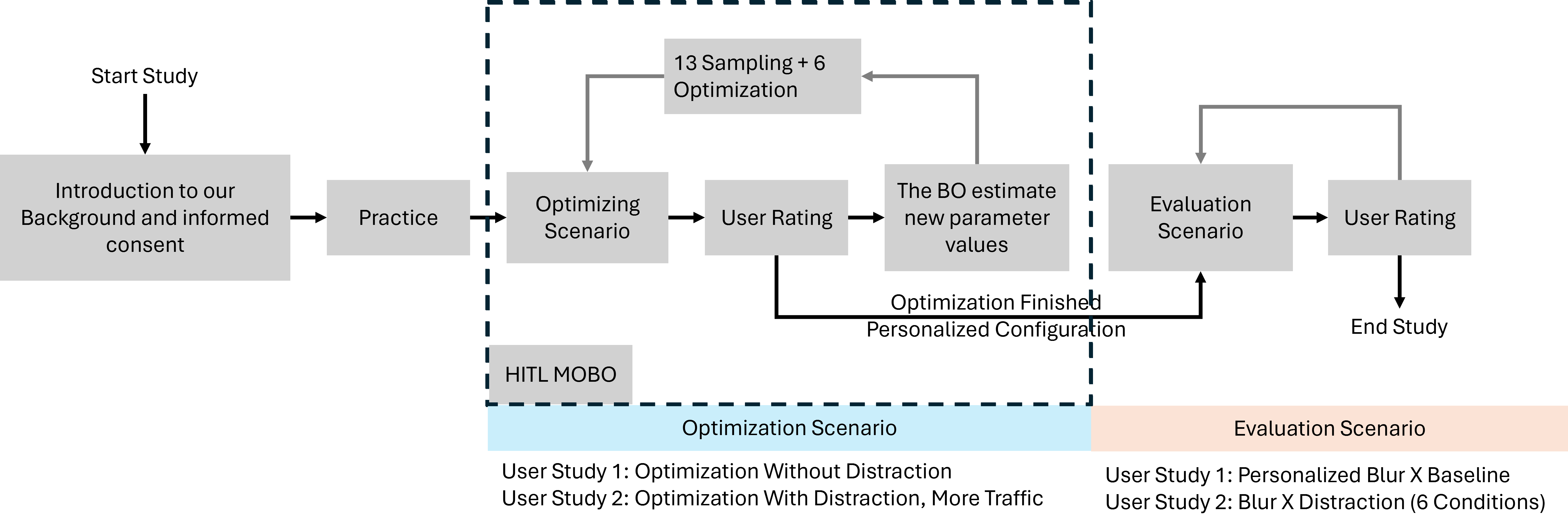}
    \caption{User study procedure, including both user study 1 and user study 2. In both user studies, participants participate in the HITL-MOBO process for personalized blur, then the main city driving scenario. Compared to user study 1, user study 2 has more traffic intensity and more conditions.}
    \label{fig:studyprocedure}
\end{figure}

After each condition, participants completed a questionnaire (see Section \ref{study1-measurements}). Finally, participants provided open-ended feedback and indicated their overall preference between the two conditions. The session lasted approximately 90 minutes, and each participant received a USD 25 Amazon gift card as compensation. Participants were informed that they could take a break or withdraw from the experiment at any time if they experienced discomfort. \autoref{fig:studyprocedure} illustrates the study procedure.

\subsection{Results of User Study 1}

For data analysis, R version 4.6.0 and RStudio version 2026.04.0 were employed. All packages were up to date in April 2026. All parametric test assumptions were tested. %For non-parametric data, we used the ARTool (ART) package by \citet{wobbrock2011art}. The test is abbreviated with ART.
We utilized the \textit{EMOA} R package~\cite{emoa} to identify the Pareto front for each participant. This front consists of all Pareto-optimal designs, each representing an optimal balance between conflicting objectives~\cite{marler_survey_2004}. 

For \autoref{fig:ps_nasa} and \autoref{fig:obj}, we used \textit{ggstatsplot}~\cite{ggstatsplot} in version 1.0.0.

\subsubsection{Blur Perceive Level}
On a scale from 1 to 7, participants clearly noticed the blurs (\m{6.13}, \sd{1.24}).

%-------------------------------------
\subsubsection{Questionnaire Ratings of the Design Objectives over Iterations} \label{res:objectives}
%-------------------------------------

\begin{figure}[ht!]
    \centering
    \small
    % Subfigure 1: All values
    \begin{subfigure}[b]{0.33\textwidth}
        \centering
        \includegraphics[width=\textwidth]{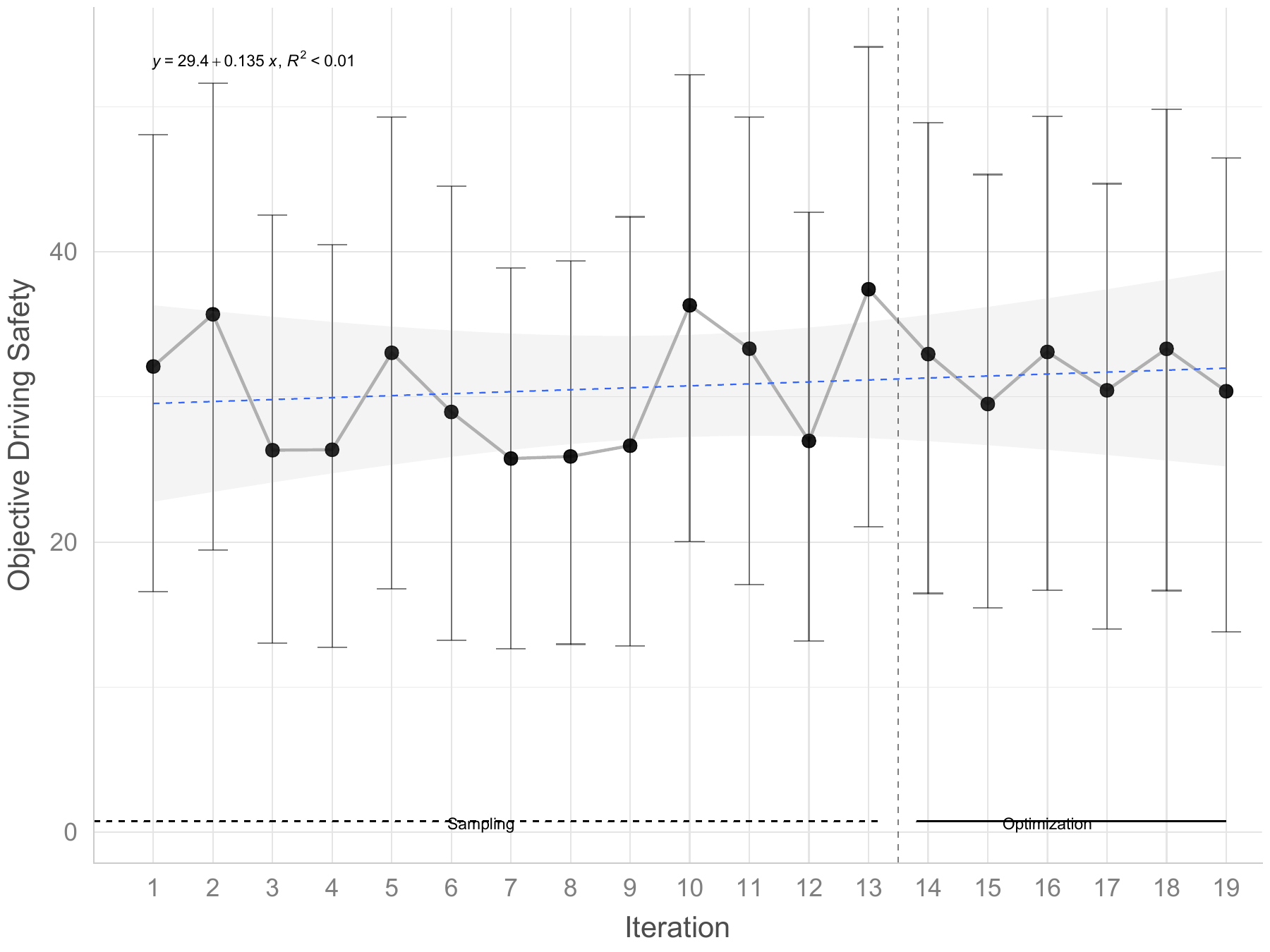}
        \caption{Progression of \textbf{Objective Driving Safety} values over iterations.}
        \label{fig:runs_errorrate}
        \Description{.}
    \end{subfigure}
    \hfill
    % Subfigure 2: MOBO values
    \begin{subfigure}[b]{0.33\textwidth}
        \centering
        \includegraphics[width=\textwidth]{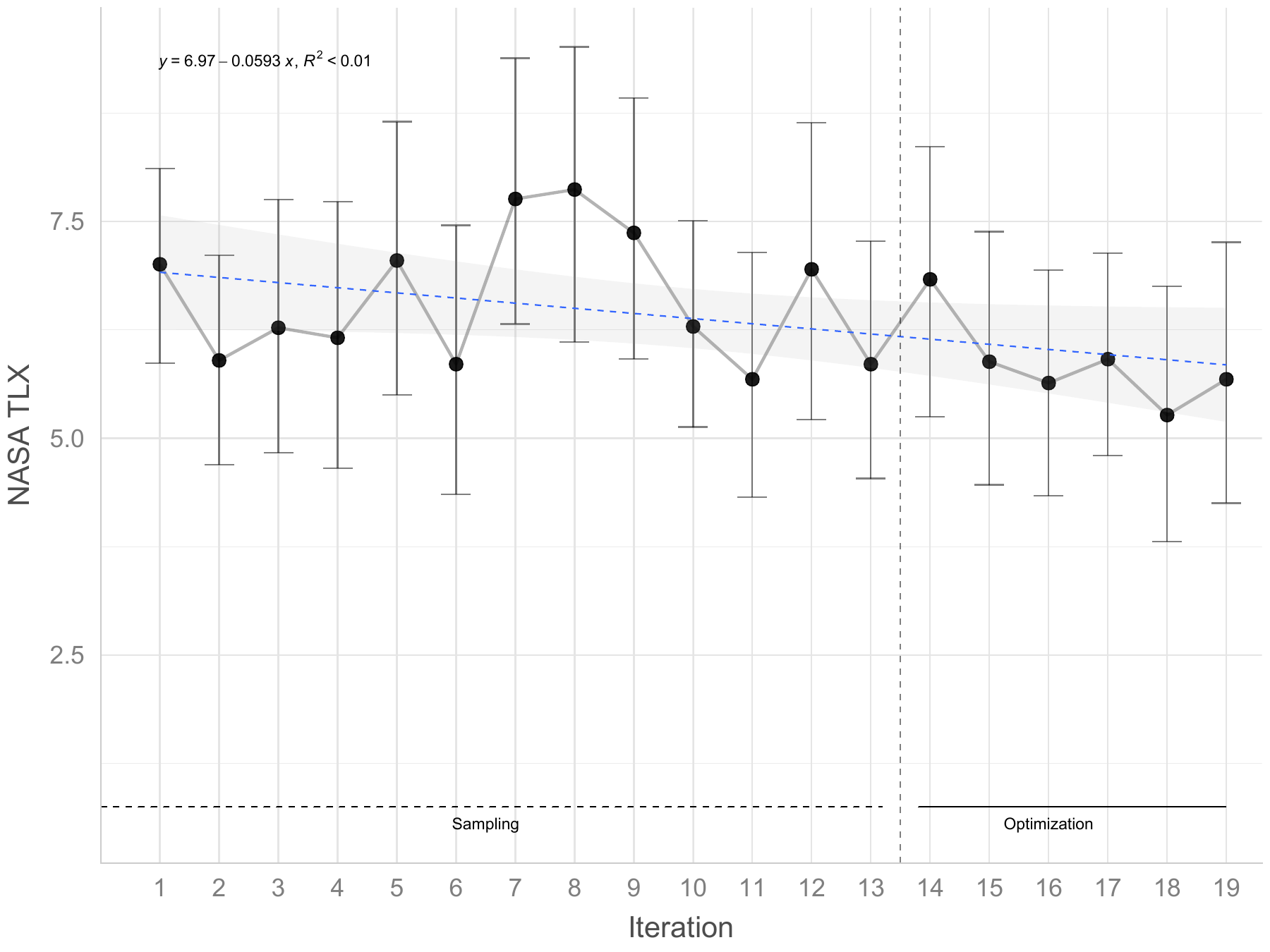}
        \caption{Progression of \textbf{NASA TLX} values over iterations.}
        \label{fig:runs_}
        \Description{}
    \end{subfigure}
        \begin{subfigure}[b]{0.33\textwidth}
        \centering
        \includegraphics[width=\textwidth]{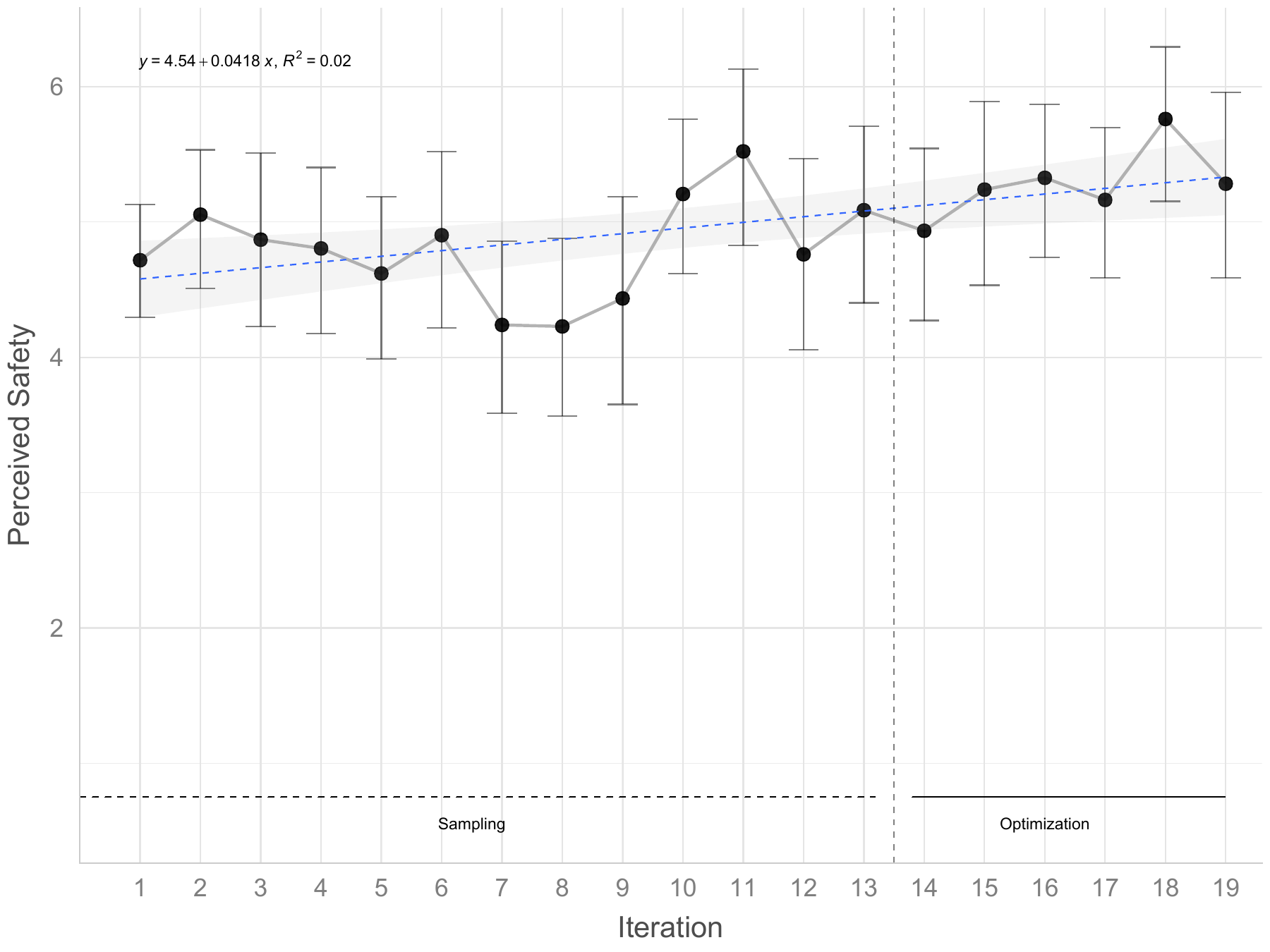}
        \caption{Progression of \textbf{Perceived Safety} values over iterations.}
        \label{fig:runs_ps}
        \Description{}
    \end{subfigure}
    \caption{User Study 1: Progression of objectives over iterations.}
    \label{fig:runs}
    \Description{}
\end{figure}

\autoref{fig:runs} shows the progression of the objectives over the HITL-MOBO iterations. Interestingly, we observe a trend in perceived safety (see \autoref{fig:runs_ps}). However, for objective driving safety and for NASA TLX, the trend is much weaker with stronger outliers (e.g., \autoref{fig:runs_errorrate} iteration 3).

%-------------------------------------
\subsubsection{Correlation between the Design Objectives} \label{res:correlation}
\begin{figure}[ht!]
    \centering
    \small
    % Subfigure 1: All values
    \begin{subfigure}[b]{0.495\textwidth}
        \centering
        \includegraphics[width=\textwidth]{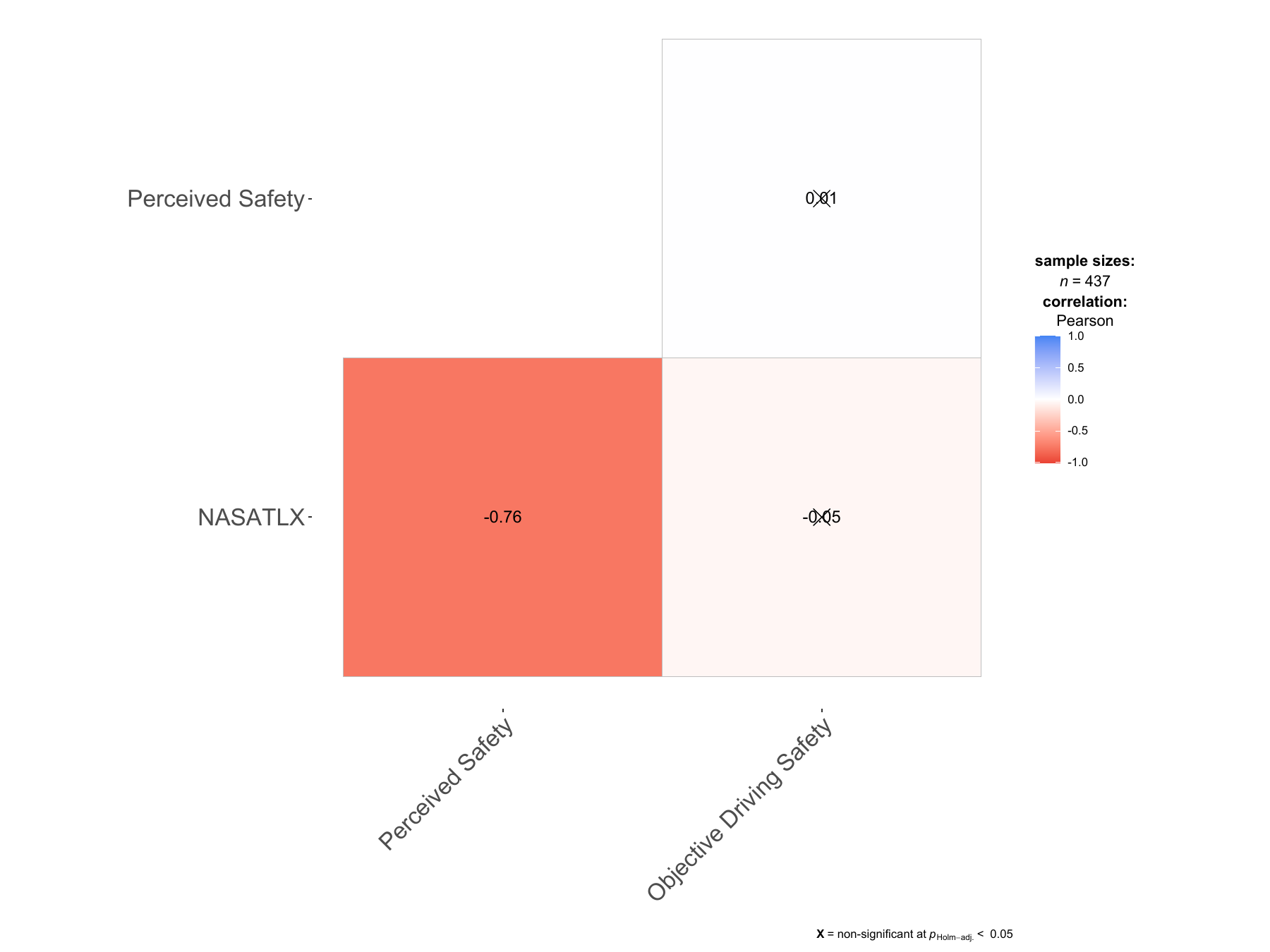}
        \caption{All objective values.}
        \label{fig:correlation_all}
        \Description{.}
    \end{subfigure}
    \hfill
    % Subfigure 2: MOBO values
    \begin{subfigure}[b]{0.495\textwidth}
        \centering
        \includegraphics[width=\textwidth]{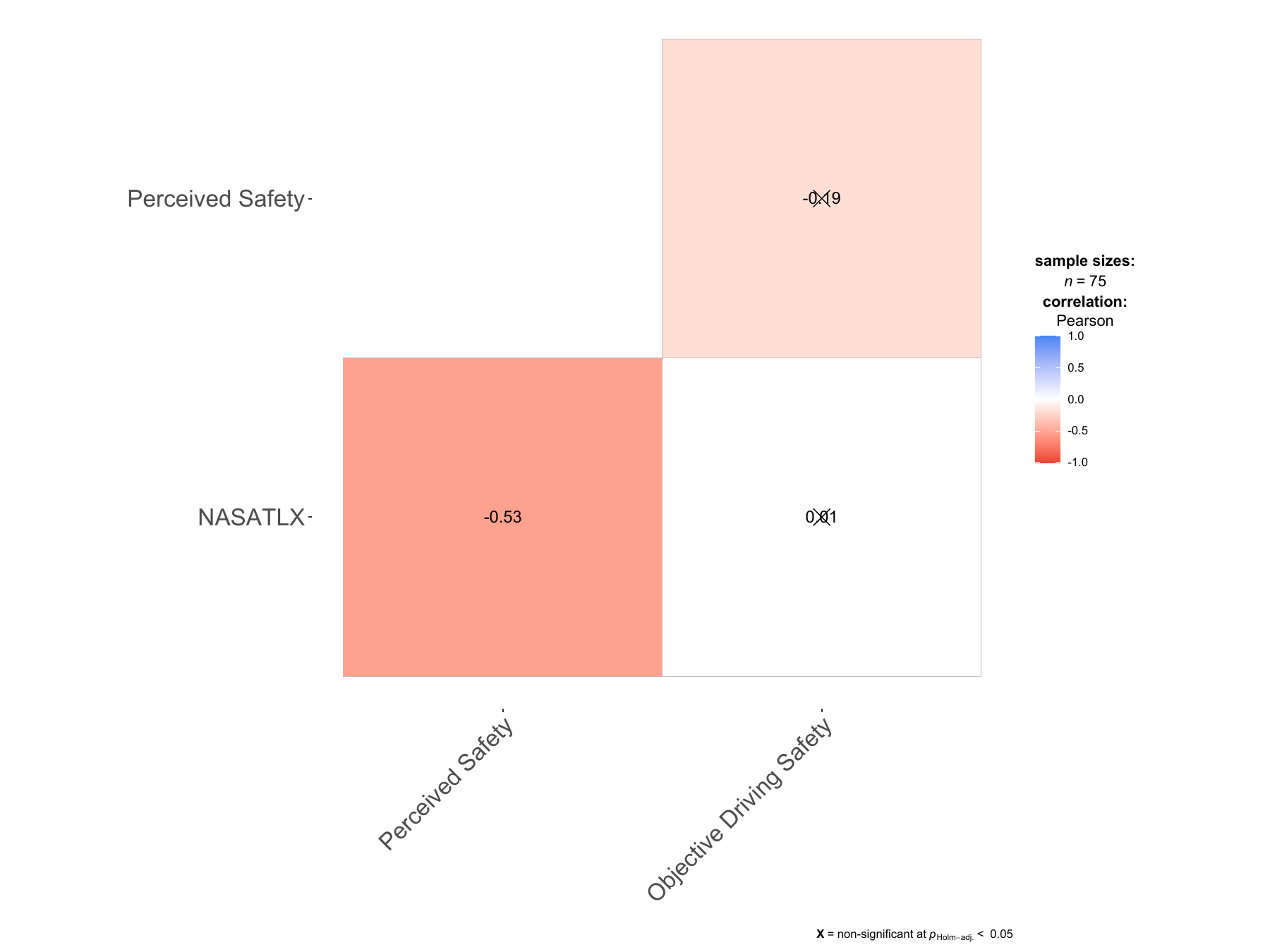}
        \caption{Only objective values for designs on the Pareto front.}
        \label{fig:correlation_mobo}
        \Description{}
    \end{subfigure}
    \caption{User Study 1: Correlation heatmaps of the design objectives. (a) includes all data points, and (b) includes only Pareto-optimal data points. ''$x$'' indicates non-significant at $p<0.05$ (adjustment: Holm).}
    \label{fig:correlation_combined}
    \Description{}
\end{figure}

%-------------------------------------
We computed Pearson correlations among objectives with Holm-adjusted p-values for \textbf{all} evaluated designs (see \autoref{fig:correlation_all}) and \textbf{Pareto‐optimal} designs only (see \autoref{fig:correlation_mobo}) to evaluate potential trade-offs between objectives, helping us understand how changes in one objective affect the others. NASA-TLX showed a strong negative correlation with Perceived Safety. However, no other correlations were found.

%-------------------------------------
\subsubsection{Design Parameter Values on the Pareto Front} \label{res:pareto-front}
%-------------------------------------

\begin{figure}[ht!]
    \centering
    \small
        \centering
        \includegraphics[width=\textwidth]{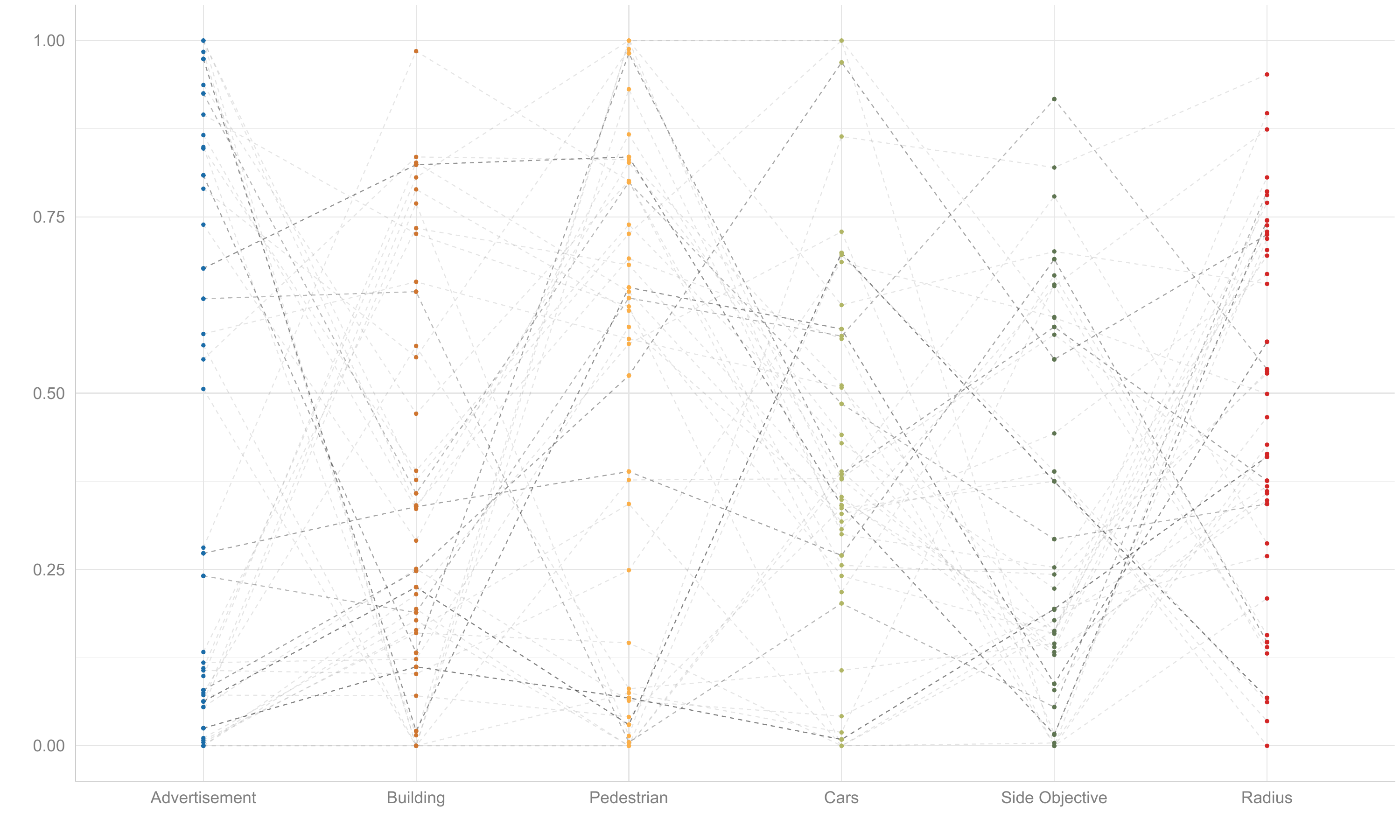}
    \caption{User Study 1: Pareto true values for all parameters.}
    \label{fig:values_all}
    \Description{}
\end{figure}

\autoref{fig:values_all} and Appendix \autoref{fig:values_all_Per_person} show the parameter values of the designs on the Pareto front. \autoref{fig:values_all} shows that the range of the blurring parameters was very diverse. 

We want to highlight that the patterns per person (see Appendix \autoref{fig:values_all_Per_person}) qualitatively differ. This indicates substantial inter-individual variation in acceptable blur configurations, but, by itself, does not show that personalization improves driving performance.

\subsubsection{Questionnaire and Objective Safety Scores} 
\label{userstudy1:result}
\begin{figure}[ht!]
    \centering
    \small
    % Subfigure 1: All values
    \begin{subfigure}[b]{0.495\textwidth}
        \centering
        \includegraphics[width=\textwidth]{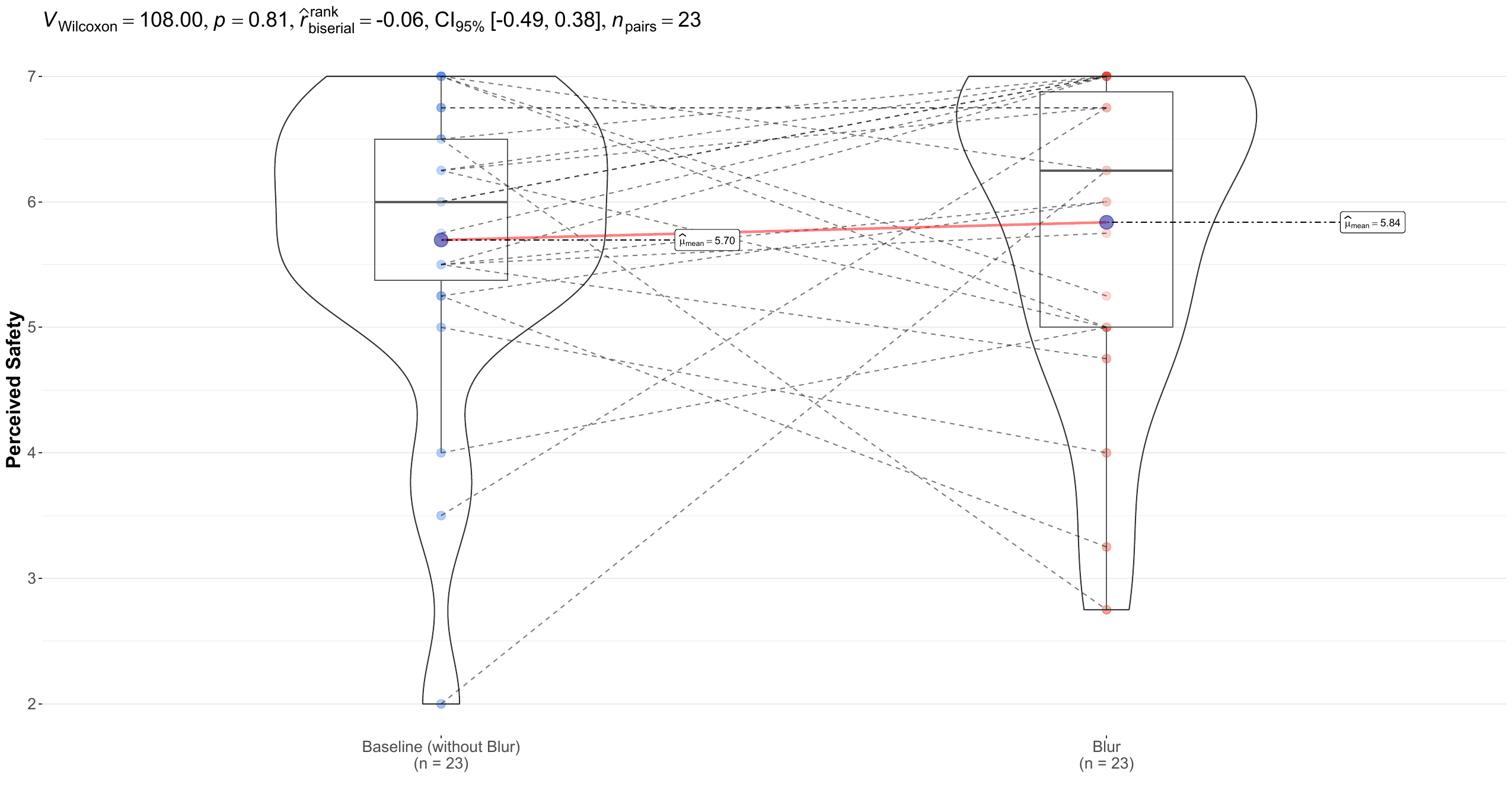}
        \caption{Perceived safety.}
        \label{fig:ps}
        \Description{.}
    \end{subfigure}
    \hfill
    % Subfigure 2: MOBO values
    \begin{subfigure}[b]{0.495\textwidth}
        \centering
        \includegraphics[width=\textwidth]{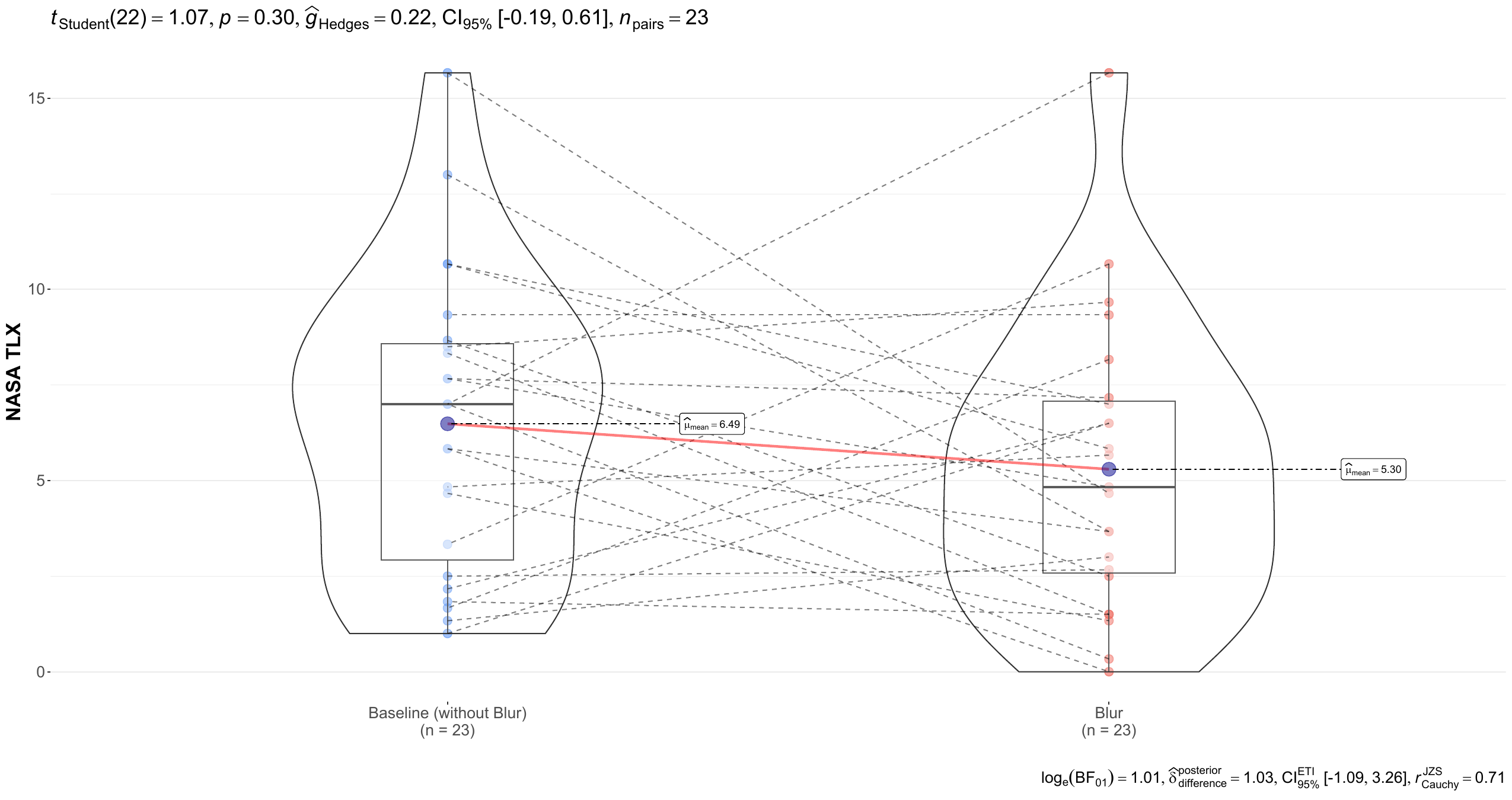}
        \caption{NASA TLX.}
        \label{fig:NASATLX}
        \Description{}
    \end{subfigure}
    \caption{User Study 1: Perceived Safety and NASA TLX. No significant difference was found.}
    \label{fig:ps_nasa}
    \Description{}
\end{figure}

\begin{figure}[ht!]
    \centering
    \small
    % Subfigure 1: All values
    \begin{subfigure}[b]{0.495\textwidth}
        \centering
        \includegraphics[width=\textwidth]{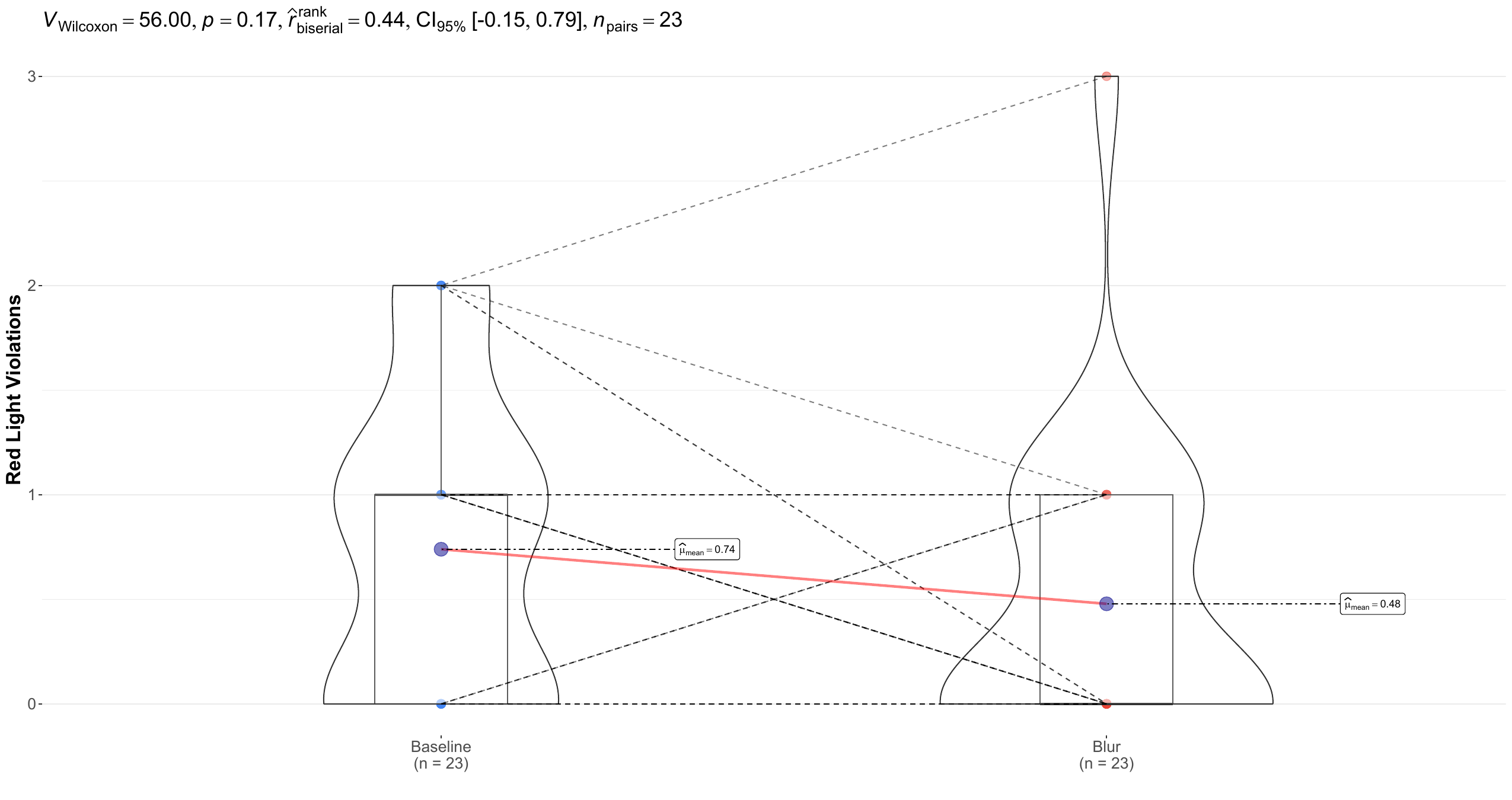}
        \caption{Red Light Violations.}
        \label{fig:RedLightViolations}
        \Description{.}
    \end{subfigure}
    \hfill
    % Subfigure 2: MOBO values
    \begin{subfigure}[b]{0.495\textwidth}
        \centering
        \includegraphics[width=\textwidth]{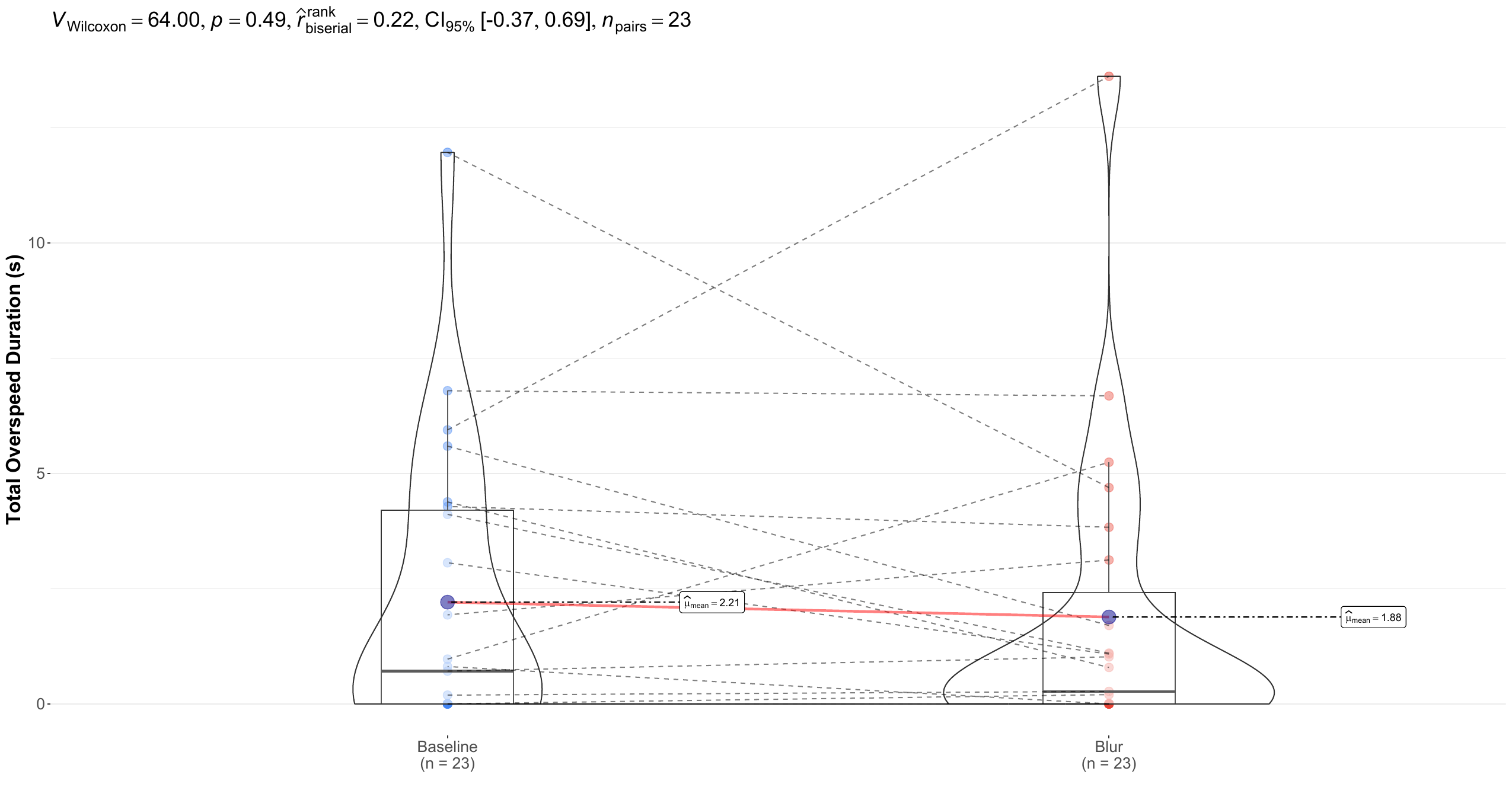}
        \caption{Total Overspeeding in seconds.}
        \label{fig:TotalOverspeedDurationSec}
        \Description{}
    \end{subfigure}
        \begin{subfigure}[b]{0.495\textwidth}
        \centering
        \includegraphics[width=\textwidth]{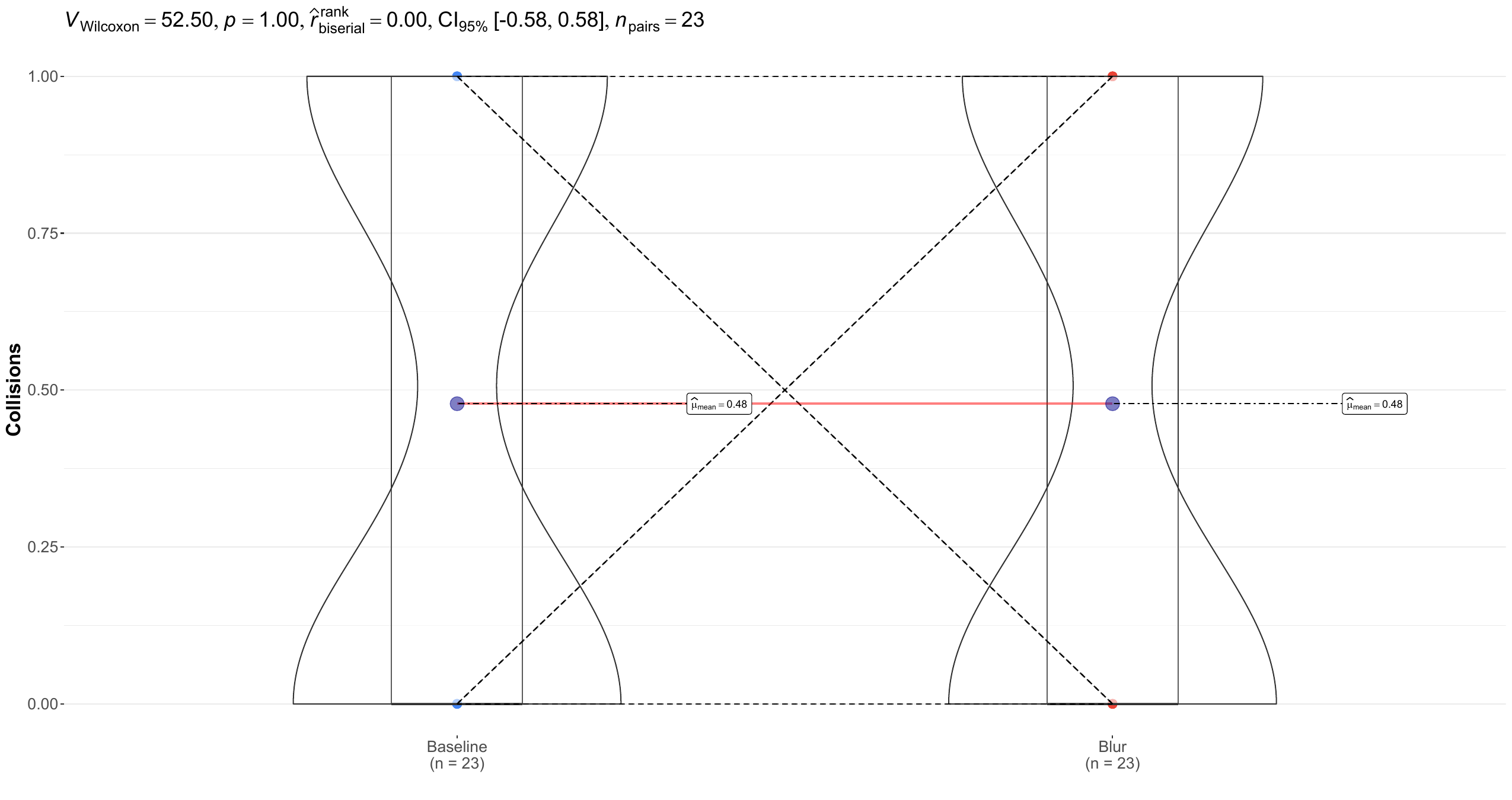}
        \caption{Collisions.}
        \label{fig:Collisions}
        \Description{.}
    \end{subfigure}
    \hfill
    % Subfigure 2: MOBO values
    \begin{subfigure}[b]{0.495\textwidth}
        \centering
        \includegraphics[width=\textwidth]{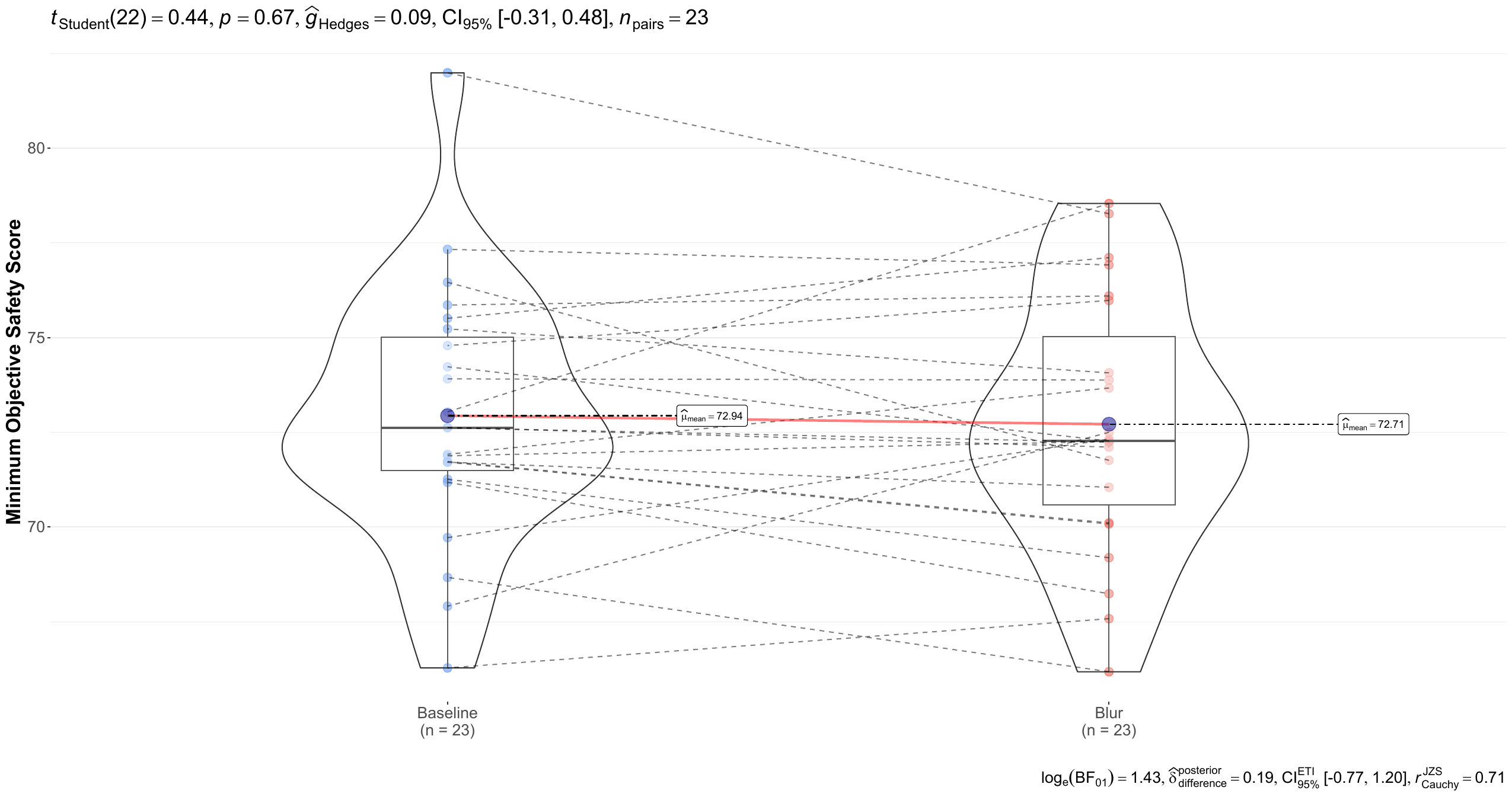}
        \caption{Minimum Objective Score.}
        \label{fig:MinSafetyScore}
        \Description{}
    \end{subfigure}
    \caption{User Study 1: Scores that contributed to the Objective Safety Score.}
    \label{fig:obj}
    \Description{}
\end{figure}

After all iterations, we compared the baseline of \textit{No Blur} with the personalized blur. We used the Pareto-front values with the highest performance (i.e., the normalized sum of all objectives). As shown in \autoref{fig:ps_nasa} and \autoref{fig:obj}, no significant differences were found for any of the objectives.

\subsubsection{Qualitative Data}
At the end of each condition, participants were invited to provide open-ended feedback, describing their impressions of the two visual conditions. After completing all conditions, participants were asked to indicate which condition they preferred and to explain the reason for their choice. They were also asked which condition they felt contributed more to driving safety. Finally, participants were given an opportunity to share general feedback on the overall experimental experience.

Because these preference judgments were collected after participants had experienced both conditions, we use them descriptively to contextualize the quantitative results. They should not be interpreted as independent predictors of blur effectiveness or as evidence that preference caused the observed subjective or behavioral responses.

\textit{Preference Condition}
Among the 23 participants, 10 preferred the Blur condition, while 13 preferred the Baseline condition.
Among those who favored Blur, seven participants mentioned that Blur helped them focus more on essential information (P1, P4, P16, P19, P20, P22, P23).
Three participants noted that blur made them feel more relaxed during driving (P8, P15, P23), for example: \textit{“I am relaxed because I do not pay attention to ads.”}

In contrast, participants who preferred the Baseline condition often felt that the blur introduced additional mental burden during driving (P2, P3, P7, P10, P17, P18). One participant stated, \textit{“I fear about the things unknown.”} (P18)
Additionally, two participants (P2, P7) explained their preference for the baseline condition as related to their long-term driving habits. For example, P7 commented:
\textit{“As someone with over four to five years of driving experience, I am used to real-world driving conditions. When I first encountered the blur, it felt somewhat unnatural. Not being able to clearly see the traffic situation made me psychologically uneasy, as I could not fully understand my surroundings.”}

\textit{Preference for Safer Driving}
Among the 23 participants, 9 preferred the Blur condition, and 14 preferred the Baseline condition for safer driving. P4 commented: \textit{"I prefer having more information regarding safety. I sometimes had to focus more to make sure I perceived a pedestrian from afar and adapt my behavior accordingly. Still, blurring made my attention definitely more focused and lightened the mental load."}

\section{User Study 2 - Blur Technique in Complex and Distracted Situation}
\label{userstudy2:overview}
The results of user study 1 showed no significant difference between the baseline and personalized blur conditions (see Section \ref{userstudy1:result}). To further investigate our RQs, we conduct user study 2 with a more complex secondary task that involves distraction.
Study 2 extends RQ1 by comparing personalized blur with predefined blur and no blur, and addresses RQ2 by testing the interaction between Blur and Distraction in a 3 $\times$ 2 factorial design. There are total 6 condition in study 2:
\begin{itemize}
    \item \textbf{No Blur (Baseline):} Participants driving without blur.
    \item \textbf{No Blur with 2-back:} Participants driving without blur under the auditory cognitive secondary task.
    \item \textbf{Defined Blur:} Participants driving with a blur profile defined by previous studies. 
    \item \textbf{Defined Blur with 2-back:} Participants driving with a blur profile defined by previous studies. With the auditory cognitive secondary task.
    \item \textbf{Personalized Blur:} Participants driving with a blur profile personalized by HITL-MOBO.
    \item \textbf{Personalized Blur with 2-back:} Participants driving with a blur profile personalized by HITL-MOBO. With the auditory cognitive secondary task.
\end{itemize}

\subsection{Scenario for User Study}
\label{userstudy2:design}

We use the same traffic rules mentioned in Section~\ref{userstudy1} and the same scenario design described in Section~\ref{userstudy1}. However, we simulate higher-intensity road traffic to investigate how drivers' performance is affected in a high-intensity, cognitively demanding scenario.

\subsection{Additional Measurements}
\label{study2-measurements}
We used the same measurements mentioned in Section~\ref{study1-measurements}. Additionally, we added the \textit{Mean Speed} and \textit{Fréchet Distance}, as the results of user study 1 suggest that our evaluation metrics lack sensitivity. We used the \textit{Fréchet Distance} to evaluate participants' driving performance, as it captures the overall similarity between the driven trajectory and a reference path by accounting for spatial alignment~\cite{Fréchet01, Fréchet02}. The \textit{Fréchet Distance} reflects the maximum deviation along the trajectory, making it particularly suitable for detecting significant deviations that may indicate safety-critical behavior \cite{Fréchet03}. The sampling rate of the Fréchet Distance is 30 Hz.

%\subsubsection{Mean Speed}
%We also calculated the change in the mean speed to evaluate blur. Mean speed was used to reflect overall driving behavior and risk perception \cite{meanspeed01}. It offers a simple yet informative indicator of caution or aggressiveness and complements metrics like lane deviation or acceleration to assess safety more comprehensively \cite{meanspeed02}.

%\subsubsection{Fréchet Distance}

\subsection{Additional Conditions and Factors}
The virtual environment remained the same as in user study 1. Additionally, we added the auditory n-back task as a distraction factor and added a defined blur as a blur factor. We conducted a 3 (Blur) $\times$ 2 (Distraction) factorial design as mentioned in \ref{userstudy2:overview}. Furthermore, we increased traffic intensity by increasing the maximum number of vehicles in the virtual environment from 40 to 80. We also updated \texttt{BoTorch}~\cite{balandat2020botorch} to version 0.16.1.

\subsubsection{Blur Factor - Defined Blur} 
Previous studies have shown that roadway information has a substantial impact on drivers’ attentional state and driving performance~\cite{advertising->distract01, advertising->distract02, advertising->distract03, advertising->distract04, sideobejct->distract01, roadcarsandroadpeople->important}. Among different types of roadside information, advertising billboards have been identified as one of the strongest sources of visual distraction, with many documented cases in which advertisements contributed to traffic accidents~\cite{advertising->distract01, advertising->distract02, advertising->distract03, advertising->distract04}. In addition, roadside elements such as buildings, sculptures, and other static objects are also considered major contributors to driver distraction, as they introduce visually salient but task-irrelevant stimuli~\cite{sideobejct->distract01}. In contrast, pedestrians and surrounding vehicles are generally regarded as safety-critical information because their behaviors are dynamic, unpredictable, and directly related to collision risk~\cite{roadcarsandroadpeople->important}. Therefore, these objects require continuous monitoring by the driver. 

Based on this distinction, in the Defined Blur condition, advertisements, buildings, and side objects are treated as blur targets and rendered with a high blur intensity, whereas pedestrians and road vehicles are treated as non-blurred objects and always remain visually clear. The \textbf{Global blur radius ($R_{\text{blur}}$)} is set to 150 meters. Given that our scenario represents urban driving, we define a \textbf{\(R_{\min}\) Safety Distance} of 40 meters, meaning that any object within this distance will not be blurred, regardless of its category, to ensure a safe braking distance and visually critical information for immediate driving decisions remains fully visible.

\subsubsection{Distracted Factor - Auditory n-back cognitive task} 
By introducing an auditory n-back task as a controlled cognitive factor in our experiment, we aim to further investigate the relationship between visual blur and cognitive workload (RQ2). We incorporated measures to prevent participant fatigue and potential learning effects from repeating the same high-difficulty task throughout the entire session. Specifically, we used a 1-back task for optimization and a 2-back task for the main evaluation.

\subsection{Participants}
Participants were recruited by the social account. We recruited 26 participants; 2 (both males) who experienced serious cybersickness were unable to complete the user study. We used data from 24 participants (16 male, 8 female, 0 non-binary) aged 24 to 39, M = 27.04 (SD = 3.21), who took part in our experiment. 20 were students, 2 researchers, and 2 employees. All participants have a valid driver’s license for M=5.51 (SD=4.36) years (ranging from 0.25 to 21 years).

\subsection{Study Procedure}
Participants were recruited via the university’s social account. Upon arrival, the instructor explained the study procedure in detail, and participants began the experiment after providing written informed consent. Then, participants had 10 minutes to practice and become familiar with the auditory n-back tasks (1-back and 2-back) and the experimental system. After practice, participants completed the optimization scenario, during which a personalized blur configuration was optimized for each individual using HITL-MOBO, while subjective questionnaire data and objective driving performance metrics were recorded. After optimization was completed, participants proceeded to the main driving task, which consisted of the 6 driving conditions mentioned above in a balanced, random sequence to remove order effects. 

Participants were instructed to drive along a fixed route in a city environment (the specific path is illustrated in \autoref{fig:task2}). Throughout the route, they encountered multiple traffic lights, pedestrians, moving vehicles, and a variety of roadside billboards. Each traffic light was accompanied by an adjacent billboard displaying different thematic content (including horror, cute, and food) to simulate realistic sources of visual distraction. After each condition, participants completed a questionnaire (see Section \ref{study2-measurements}). 

Finally, participants provided open-ended feedback and indicated their overall preference across the 3 visual conditions. The session lasted approximately 120 minutes, and each participant received a USD 30 Amazon gift card as compensation. Participants were informed that they could take a break or withdraw from the experiment at any time if they experienced discomfort. The procedure of user study 2 is similar to that of user study 1 (\autoref{fig:studyprocedure}), but using a different scenario setup and configurations (see Section \ref{userstudy2:design}).

\subsection{Results of User Study 2}
We used the same data analysis approach as in study 1.

%\paragraph{Data Analysis}

% For \autoref{fig:ps_nasa} and \autoref{fig:obj}, we used \textit{ggstatsplot}~\cite{ggstatsplot} in version 0.13.4.

%-------------------------------------
\subsubsection{Blur Perceive Level}
On a scale from 1 to 7, participants clearly noticed the blurs (\m{5.88}, \sd{1.51}).

\subsubsection{Questionnaire Ratings of the Design Objectives over Iterations}

% \autoref{fig:study2runs} shows the progression of the three optimization objectives across the 19 HITL-MOBO iterations in User Study 2. Objective Driving Safety and NASA-TLX showed only weak descriptive trends over iterations (see \autoref{fig:study2_runs_errorrate} and \autoref{fig:study2_runs_}), while Perceived Safety remained largely stable in the distracted scenario (see \autoref{fig:study2_runs_ps}).

\begin{figure}[ht!]
    \centering
    \small
    % Subfigure 1: All values
    \begin{subfigure}[b]{0.33\textwidth}
        \centering
        \includegraphics[width=\textwidth]{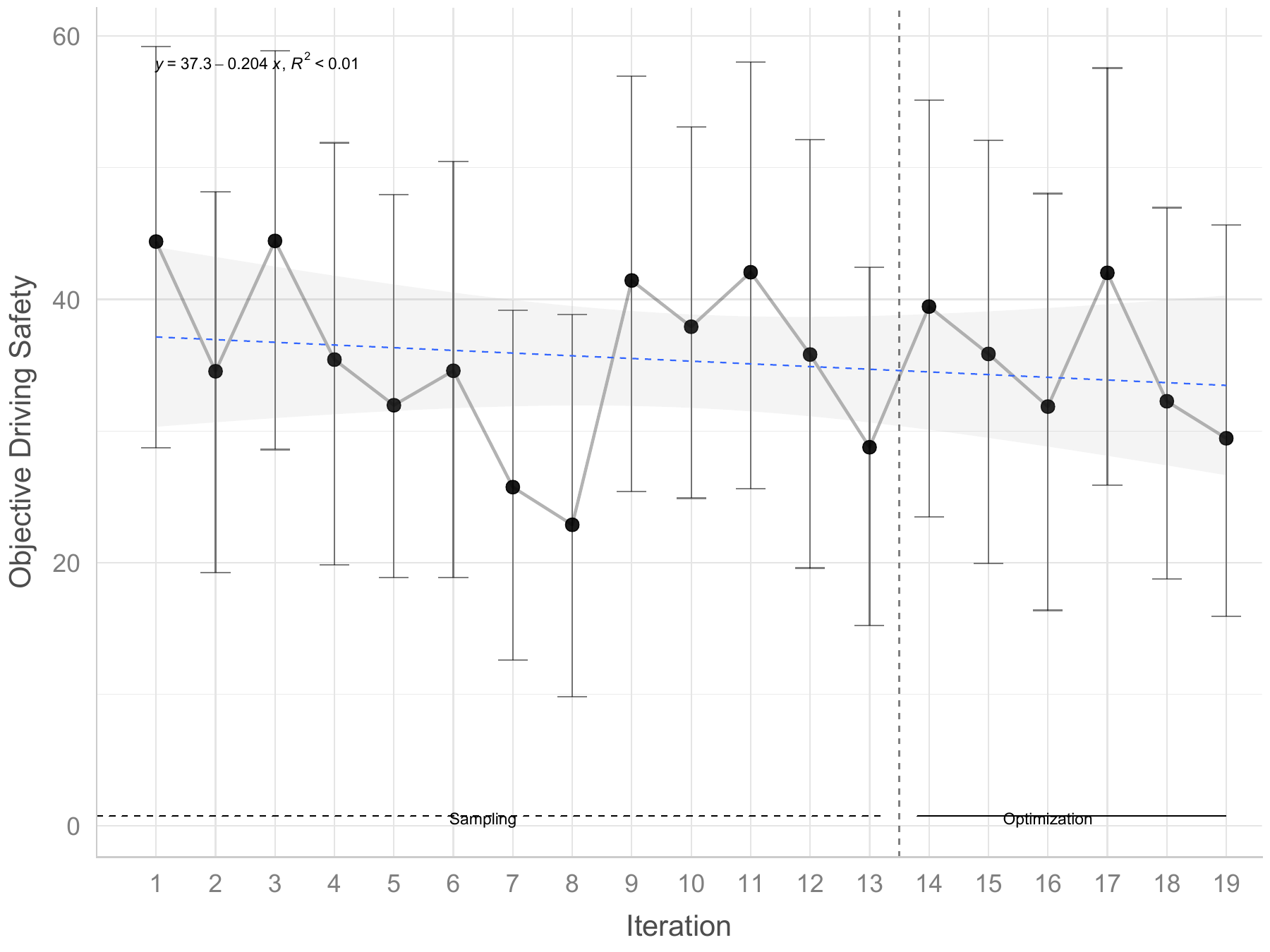}
        \caption{Progression of \textbf{Objective Driving Safety} values over iterations.}
        \label{fig:study2_runs_errorrate}
        \Description{.}
    \end{subfigure}
    \hfill
    % Subfigure 2: MOBO values
    \begin{subfigure}[b]{0.33\textwidth}
        \centering
        \includegraphics[width=\textwidth]{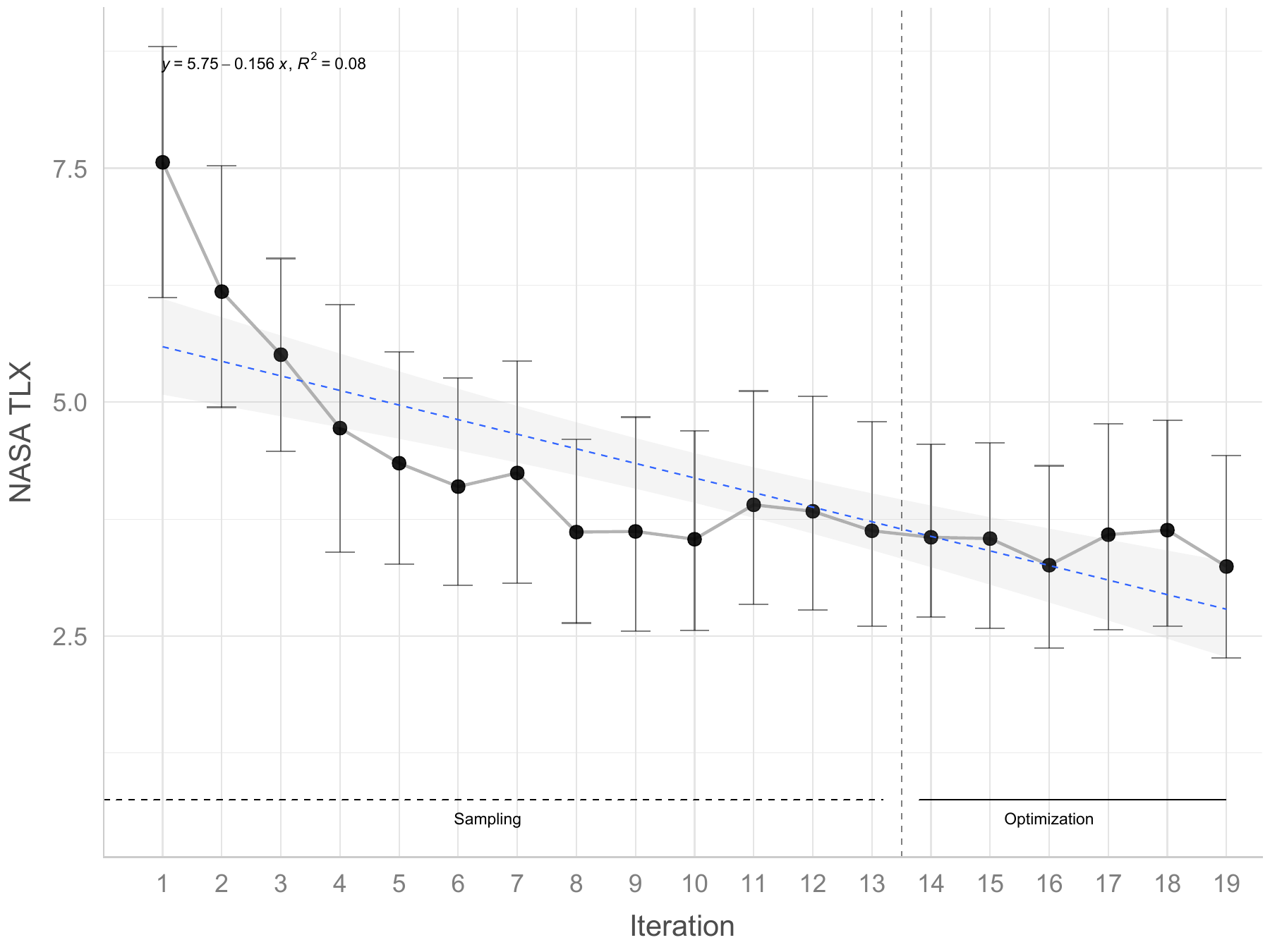}
        \caption{Progression of \textbf{NASA TLX} values over iterations.}
        \label{fig:study2_runs_}
        \Description{}
    \end{subfigure}
        \begin{subfigure}[b]{0.33\textwidth}
        \centering
        \includegraphics[width=\textwidth]{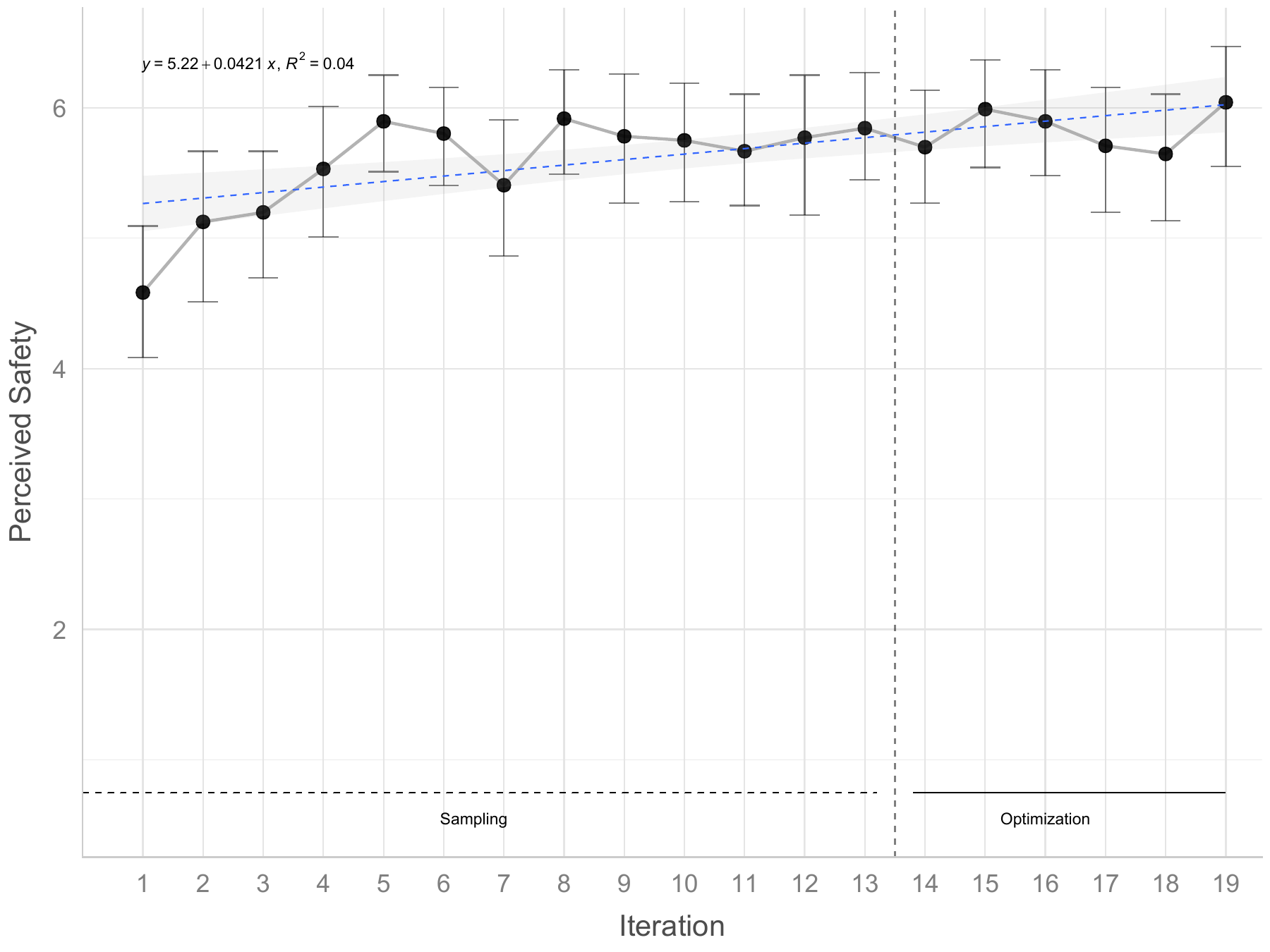}
        \caption{Progression of \textbf{Perceived Safety} values over iterations.}
        \label{fig:study2_runs_ps}
        \Description{}
    \end{subfigure}
    \caption{User Study 2: Progression of objectives over iterations.}
    \label{fig:study2runs}
    \Description{}
\end{figure}

\autoref{fig:study2runs} shows the progression of the three optimization objectives across the 19 HITL-MOBO iterations in User Study 2. Perceived Safety showed a modest overall increase, whereas NASA-TLX decreased primarily during the early sampling iterations and subsequently stabilized during the optimization phase (see \autoref{fig:study2_runs_ps} and \autoref{fig:study2_runs_}). In contrast, Objective Driving Safety remained highly variable and showed no clear improvement across iterations (see \autoref{fig:study2_runs_errorrate}). Given the low $R^2$ values, these patterns should be interpreted descriptively and may partly reflect task familiarization or adaptation rather than improvements specifically attributable to the MOBO optimization phase.

%-------------------------------------

\subsubsection{Correlation between the Design Objectives.}
When considering all sampled configurations, perceived safety is strongly driven by cognitive workload, while remaining largely decoupled from objective driving safety. Importantly, this dissociation persists even when restricting the analysis to Pareto-optimal solutions, indicating that the MOBO preserves meaningful trade-offs rather than collapsing subjective and objective safety into a single dimension.

\begin{figure}[ht!]
    \centering
    \small
    % Subfigure 1: All values
    \begin{subfigure}[b]{0.495\textwidth}
        \centering
        \includegraphics[width=\textwidth]{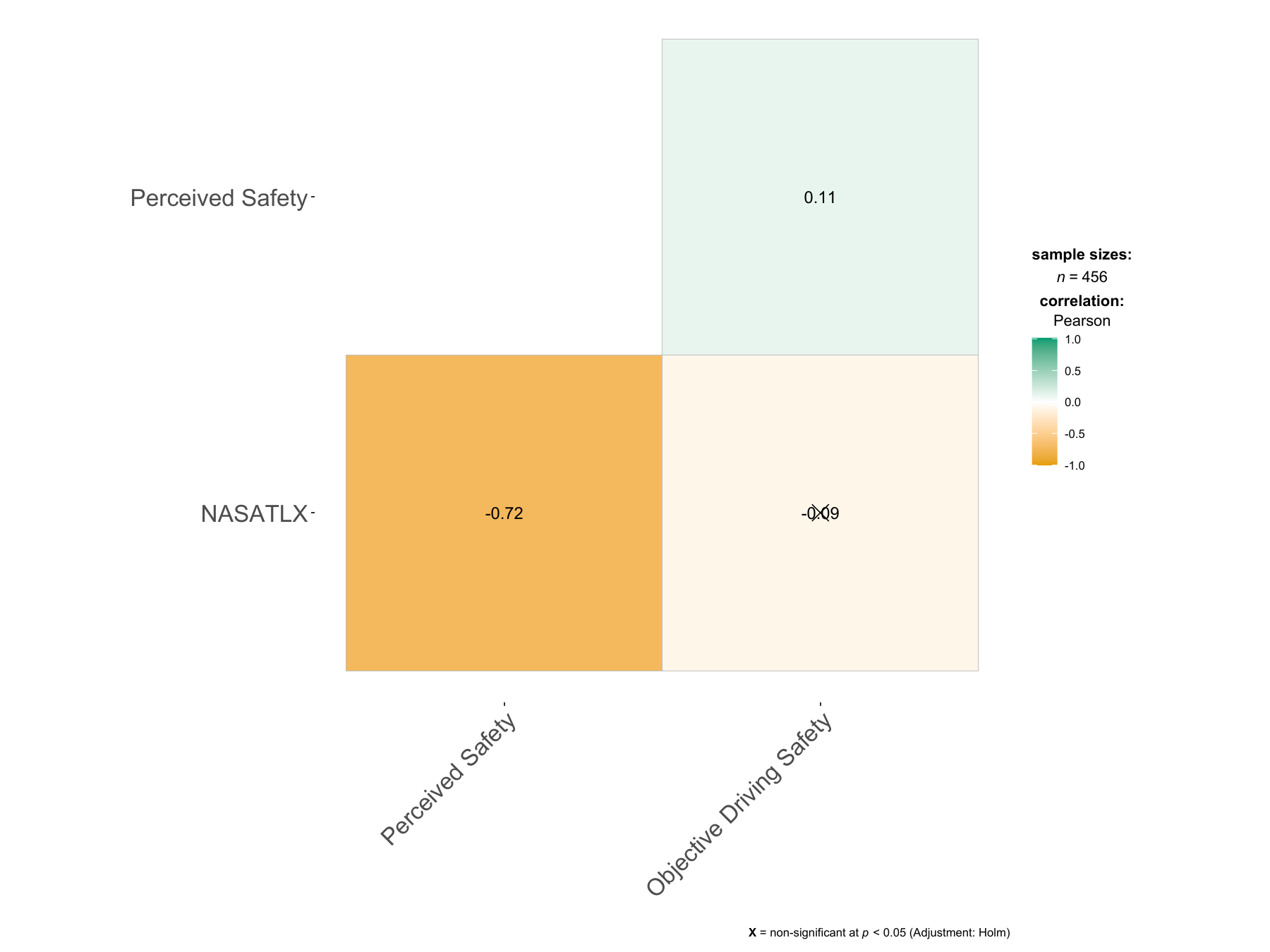}
        \caption{All objective values.}
        \label{fig:study2_correlation_all}
        \Description{.}
    \end{subfigure}
    \hfill
    % Subfigure 2: MOBO values
    \begin{subfigure}[b]{0.495\textwidth}
        \centering
        \includegraphics[width=\textwidth]{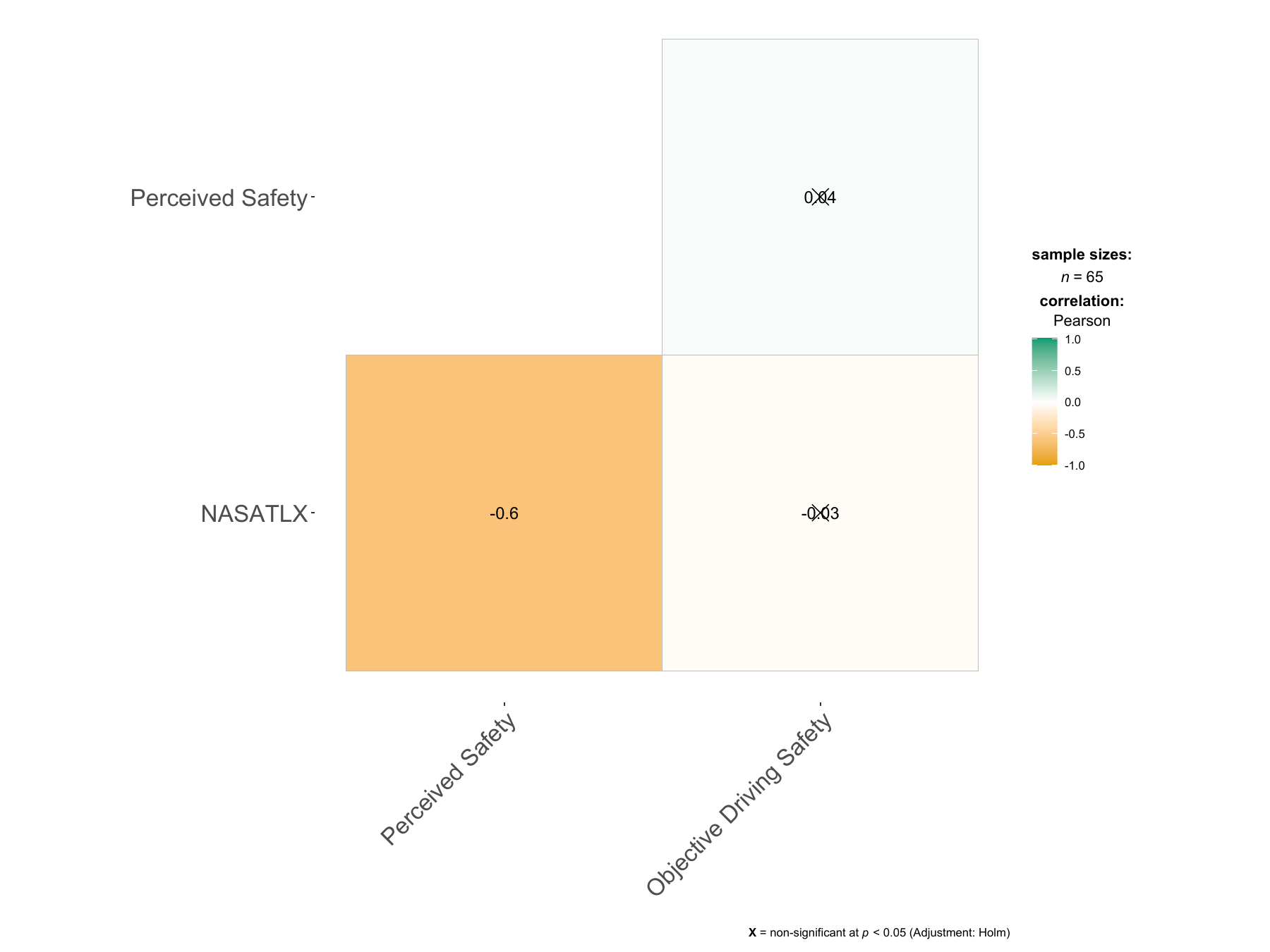}
        \caption{Only objective values for designs on the Pareto front.}
        \label{fig:study2_correlation_mobo}
        \Description{}
    \end{subfigure}
    \caption{User Study 2: Correlation heatmaps of the design objectives. (a) includes all data points, and (b) includes only Pareto-optimal data points. ''$x$'' indicates non-significant at $p<0.05$ (adjustment: Holm).}
    \label{fig:study2_correlation_combined}
    \Description{}
\end{figure}

%-------------------------------------
\subsubsection{Design Parameter Values on the Pareto Front} 

\begin{figure}[ht!]
    \centering
    \small
        \centering
        \includegraphics[width=\textwidth]{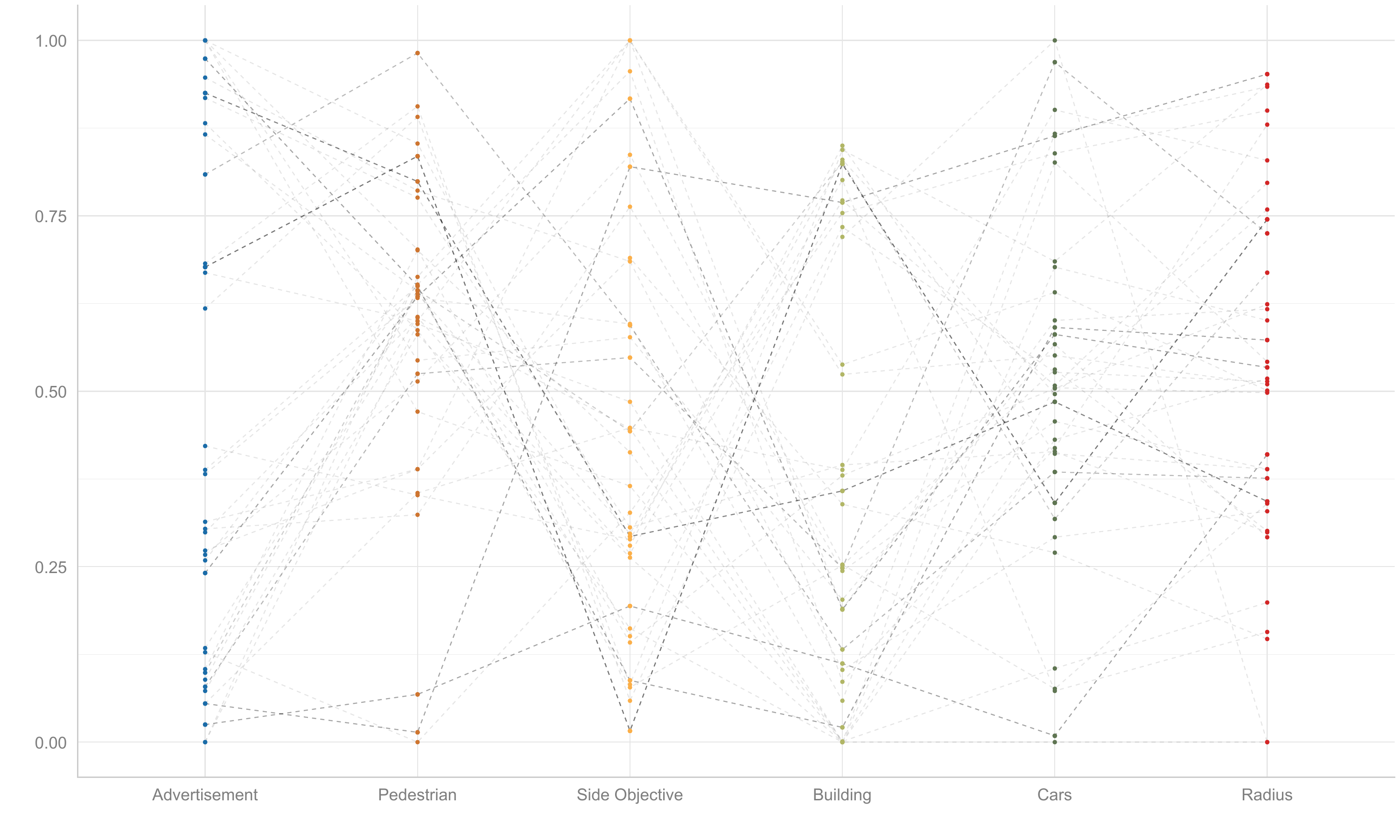}
    \caption{User Study 2: Pareto true values for all parameters.}
    \label{fig:study2_values_all}
    \Description{}
\end{figure}

\autoref{fig:study2_values_all} and Appendix \autoref{fig:study2_values_all_Per_person} show the parameter values of the designs on the Pareto front. \autoref{fig:study2_values_all} shows that the range of the blurring parameters was very diverse. 

This indicates substantial inter-individual variation in acceptable blur configurations, but, by itself, does not show that personalization improves driving performance.

%-------------------------------------

\subsubsection{Main Driving Task Questionnaire.}
\paragraph{Perceived Safety}
The ART found a significant main effect of Distraction on perceived safety (\F{1}{23}{23.47}, \pminor{0.001}, $\eta_{p}^{2}$ = 0.51, 95\% CI: [0.25, 1.00]). No significant main effect on perceived safety of Blur and no interaction between Blur and Distraction were observed. With the 2-back task, perceived safety was significantly lower.

\paragraph{NASA-TLX and Subscales of NASA-TLX}
The ART found a significant main effect of Distraction on NASATLX (\F{1}{23}{33.30}, \pminor{0.001}, $\eta_{p}^{2}$ = 0.51, 95\% CI: [0.25, 1.00]). With the 2-back task, NASA-TLX was significantly higher.

The ART found a significant main effect of Blur on Mental Demand (\F{2}{46}{4.58}, \p{0.015}, $\eta_{p}^{2}$ = 0.17, 95\% CI: [0.02, 1.00]). The ART found a significant main effect of Distraction on Mental Demand (\F{1}{23}{31.43}, \pminor{0.001}, $\eta_{p}^{2}$ = 0.58, 95\% CI: [0.34, 1.00]). However, a post hoc test found no significant differences in Mental Demand. The ART revealed a significant main effect of Distraction across multiple NASA-TLX subscales, including Physical Demand (\F{1}{23}{20.54}, \pminor{0.001}, $\eta_{p}^{2}$ = 0.47, 95\% CI [0.22, 1.00]), Temporal Demand (\F{1}{23}{25.01}, \pminor{0.001}, $\eta_{p}^{2}$ = 0.52, 95\% CI [0.27, 1.00]), Performance (\F{1}{23}{15.25}, \pminor{0.001}, $\eta_{p}^{2}$ = 0.40, 95\% CI [0.14, 1.00]), Effort (\F{1}{23}{23.45}, \pminor{0.001}, $\eta_{p}^{2}$ = 0.50, 95\% CI [0.25, 1.00]), and Frustration (\F{1}{23}{27.61}, \pminor{0.001}, $\eta_{p}^{2}$ = 0.55, 95\% CI [0.30, 1.00]).

\subsubsection{Main Driving Task Error Rate.}
The ART found no significant effects on Collisions, Red Light Violations, Safety Score, and Total Overspeed Duration.

\subsubsection{Fréchet Distance and Mean Speed}
The ART found no significant effects on the Fréchet Distance and Mean Speed.

\begin{table*}[t]
\centering
\caption{Summary of ART ANOVA results across subjective and objective measures in User Study 2.
F-values are reported with corresponding p-values. Significant effects ($p < .05$) are shown in bold.}
\label{tab:anova_summary_restructured}
\begin{tabular}{lcccccc}
\toprule
\textbf{Measure} &
\multicolumn{2}{c}{\textbf{Blur}} &
\multicolumn{2}{c}{\textbf{Distraction}} &
\multicolumn{2}{c}{\textbf{Blur $\times$ Distraction}} \\
\cmidrule(lr){2-3}
\cmidrule(lr){4-5}
\cmidrule(lr){6-7}
 & F & p & F & p & F & p \\
\midrule

% ======================
\multicolumn{7}{l}{\textbf{Subjective Measures}} \\
\midrule

Perceived Safety
& 0.61 & .550
& \textbf{23.47} & \textbf{<.001}
& 0.42 & .661 \\

NASA-TLX (RAW TLX Score)
& 2.09 & .136
& \textbf{33.30} & \textbf{<.001}
& 1.03 & .367 \\

\quad Mental Demand
& \textbf{4.58} & \textbf{.015}
& \textbf{31.43} & \textbf{<.001}
& 2.01 & .146 \\

\quad Physical Demand
& 0.78 & .465
& \textbf{20.54} & \textbf{<.001}
& 2.85 & .068 \\

\quad Temporal Demand
& 0.93 & .400
& \textbf{25.01} & \textbf{<.001}
& 0.95 & .396 \\

\quad Performance
& 1.60 & .212
& \textbf{15.25} & \textbf{<.001}
& 0.84 & .438 \\

\quad Effort
& 1.09 & .346
& \textbf{23.45} & \textbf{<.001}
& 0.90 & .414 \\

\quad Frustration
& 2.17 & .126
& \textbf{27.61} & \textbf{<.001}
& 0.51 & .602 \\

\midrule
% ======================
\multicolumn{7}{l}{\textbf{Objective Measures}} \\
\midrule

Mean Speed (km/h)
& 0.70 & .500
& 0.00 & .963
& 1.42 & .251 \\

Fréchet Distance (3D)
& 0.26 & .775
& 0.29 & .593
& 0.28 & .761 \\

Collisions
& 0.19 & .829
& 0.33 & .570
& 0.06 & .938 \\

Red Light Violations
& 2.37 & .105
& 0.03 & .857
& 3.02 & .059 \\

Overspeed Duration (s)
& 0.37 & .696
& 0.26 & .615
& 0.20 & .817 \\

Min Safety Score
& 1.68 & .197
& 0.24 & .631
& 0.13 & .882 \\

\bottomrule
\end{tabular}
\end{table*}

\subsubsection{Qualitative Data}

Qualitative data in User Study 2 were collected using the same procedure as in User Study 1. 
Again, preference judgments were collected after participants had experienced the experimental conditions. We therefore use these responses descriptively to contextualize participants' interpretations of blur, rather than as causal evidence that preference determined the effects of blur on workload, perceived safety, or driving performance.

For \textit{Preference Condition}, 7 of the 24 valid participants preferred a blur condition, all of whom selected personalized blur, whereas 17 preferred the baseline condition. For \textit{Preference for Safer Driving}, 7 participants preferred a blur condition (1 defined blur and 6 personalized blur), whereas 17 preferred the baseline condition.

Participants who preferred blur described it as a useful attentional filter. They reported that blur helped them focus on driving-relevant information by reducing the salience of irrelevant visual details. For example, P5 stated that “blur freed my attention from irrelevant details,” and P11 described blur as having “reduced distraction without affecting safety.” In contrast, participants who preferred the baseline condition often associated blur with fatigue, perceptual discomfort, or reduced confidence in their visual judgment, particularly when driving under secondary-task load. Several participants reported dizziness or visual instability, for example, “Blur makes me feel dizzy” (P6), “Blur increased fatigue” (P9), and “Blur affects my judgment” (P1). These comments suggest that blur can become a source of subjective discomfort when cognitive demands are already high.

Across participants, three recurring themes emerged. First, blur produced mixed perceptual experiences. Some participants experienced blur as a reduction of visual distraction, whereas others reported visual fatigue, dizziness, or impaired perceptual clarity. Second, blur could either reduce or create a distraction. Although the technique was intended to suppress irrelevant information, some participants reported that blurred content became more salient because they tried to infer what was hidden; as P15 noted, “Blur made me pay more attention to what the content actually was.” Third, cognitive load shaped how blur was noticed and experienced. Under the 2-back condition, some participants barely noticed the blur, for example, “I almost did not notice whether it was blurred” (P16), whereas others described the combination of driving and the secondary task as overwhelming; P14 stated, “My brain could not keep up.” This indicates that secondary-task load can dominate the driving experience and amplify fatigue-related responses to blur.

Overall, the qualitative findings show that blur is not a uniformly beneficial visual simplification technique. It can support attentional focus for some users, but it can also increase uncertainty, fatigue, or discomfort for others, especially during multitasking.

\section{Discussion}

This study explored how controllable blur techniques affect driving behavior with two user studies in a simulated VR urban driving scenario. In the first study, we evaluated how the personalized blur affects drivers' performance. In the second one, we explore further whether personalization is a necessary and distracting factor under higher traffic intensity.
Our findings investigated both research questions and provided design and methodological insights for future in-vehicle interface development.

\subsection{How Blur Techniques Impact Drivers' Performance}
Across both studies, blur did not produce reliable group-level improvements in objective driving performance relative to baseline. This suggests that blur does not act as a general-purpose performance aid in driving. However, it changes the perceptual conditions under which drivers allocate attention. The qualitative feedback helps explain why this effect did not translate into consistent improvements in design objectives: blur appears to create two competing mechanisms that can cancel each other out at the aggregate level.

First, several participants described blur as a perceptual filter that reduced attentional capture by task-irrelevant content, such as advertisements and visually dense environmental textures. For these participants, blur supported a strategy of prioritizing lane keeping, traffic signals, and dynamic hazards. This is consistent with the view of distraction as a dynamic process of engagement and disengagement, where interventions may alter attentional allocation without necessarily producing stable improvements in design objectives across scenarios \cite{lee2014dynamics}. In this interpretation, blur can help by suppressing low-relevance visual detail without reducing drivers' confidence in interpreting the driving scene.

Second, other participants reported that blur increased perceptual uncertainty and visual effort, as reflected in reduced confidence, fatigue, dizziness, or discomfort. In these cases, blur did not remove distraction; it became an additional object of interpretation. Participants sometimes appeared to monitor what was blurred, judge whether the blurred content was relevant, or compensate for reduced visual clarity. This interpretation is consistent with evidence that drivers can miscalibrate subjective judgments of risk and performance relative to objective driving quality \cite{charlton2014risk, verster2012predict}. Blur may therefore improve comfort or focus for some participants while increasing monitoring demands for others, yielding heterogeneous responses without a consistent mean effect on driving performance.

This heterogeneity may also be related to individual differences in visual acuity and sensitivity to blur. For example, participants with myopia or those accustomed to slightly blurred vision may be less sensitive to our blur manipulation \cite{myopiaBlur, myopiaBlur2}. Conversely, participants who rely heavily on high-resolution peripheral or environmental cues may experience the same manipulation as a loss of control or a reduction in scene legibility. As a result, the same blur setting can be highly noticeable for some drivers, barely perceptible for others, and actively disruptive for another subgroup. This limits the effectiveness of blur as a uniform visual guidance technique across users.

Finally, the reported dizziness and fatigue in the more demanding conditions are compatible with known individual differences in VR discomfort and with mechanisms of visual fatigue driven by degraded or conflicting focus and depth cues \cite{rebenitsch2016cybersickness, hoffman2008vergence, shibata2011zone}. We did not measure simulator sickness with a standardized questionnaire, so these reports should be interpreted as self-reported discomfort rather than validated cybersickness outcomes. Nevertheless, they strengthen the central interpretation of our results: blur should not be treated as a default interface state for driving. It is better understood as a conditional aid whose benefits depend on task demand, object relevance, and individual tolerance for visual uncertainty, arguing for personalized blur techniques.

\subsection{Implications of the Pareto Front Parameter Diversity}
The Pareto front analyses in both studies (User Study 1: \autoref{fig:values_all}, \autoref{fig:values_all_Per_person}; User Study 2: \autoref{fig:study2_values_all}, and \autoref{fig:study2_values_all_Per_person}) show wide dispersion in non-dominated parameter settings. This dispersion indicates that there is no single blur configuration that jointly optimizes our objectives for all drivers. Instead, participants occupy distinct regions of the trade-off surface, reflecting heterogeneous tolerance for uncertainty, different preferences for where visual detail should be preserved, and different responses to visually degraded information.

However, diverse Pareto-optimal designs should not be interpreted as evidence that personalization necessarily improves objective driving performance. Pareto optimality is defined only with respect to the selected objectives, the sampled design space, and the sensitivity of the corresponding measurements. In our data, subjective objectives were more strongly correlated with each other than with objective driving safety. This indicates partial misalignment between what feels safe or manageable and what yields measurable driving-performance benefits in the tested scenarios \cite{charlton2014risk, verster2012predict}. The Pareto front, therefore, primarily functions as a diagnostic tool. It reveals user-specific trade-offs that are hidden by averaged comparisons, but it does not, by itself, validate blur as an effective safety intervention.

This distinction is important for interpreting HITL-MOBO in safety-critical contexts. The optimizer can identify configurations that are non-dominated with respect to the measured objectives, but it cannot guarantee that these configurations improve actual driving safety if the objective functions capture only a partial representation of safety-relevant behavior. The broad parameter diversity, therefore, supports the need for personalization at the level of perception and comfort, while the null driving performance effects show that personalization alone is insufficient as evidence of safety benefit.

\subsection{Is HITL-MOBO Sufficient for Safety-Critical Adaptation?}
HITL-MOBO is well-suited for efficiently exploring a multi-parameter design space with sparse and noisy human feedback. In our studies, it helped map individual trade-offs and identify user-specific blur configurations with relatively few feedback iterations. The comparison with defined blur and baseline conditions, however, did not show a robust advantage of HITL-MOBO for objective driving performance. Its value in this work is therefore diagnostic rather than performance-improving. It reveals how users differ in their tolerance for blur and how subjective preferences can diverge from objective safety-related outcomes.

This limitation matters because driving risk evolves continuously and often requires rapid re-engagement. Post-trial ratings are temporally coarse, and they can prioritize comfort, perceived clarity, or subjective control over context-appropriate safety behavior. Inattention and delayed responses are strongly associated with crash and near-crash risk in naturalistic driving \cite{klauer2006inattention}, and distraction is commonly characterized as a dynamic engagement process rather than a single event \cite{lee2014dynamics}. A system that optimizes only from post-trial preference or perceived safety can therefore select configurations that are acceptable in routine segments but inappropriate under high demand, especially for users who experience blur-induced uncertainty, fatigue, or discomfort \cite{rebenitsch2016cybersickness}.

These findings do not argue against HITL-MOBO for adaptive interfaces. Rather, they clarify the boundary conditions under which it should be used. HITL-MOBO can support personalization when the goal is to learn acceptable perceptual settings, but safety-critical adaptation requires additional online constraints. A practical direction is a hybrid approach: use HITL-MOBO to learn an individual default blur profile and acceptable parameter bounds, then gate or modulate blur online using objective risk proxies and driver-state proxies. As a simple baseline, a rule-based controller could reduce or disable blur when elevated risk is detected, when safety-critical objects enter the scene, or when signs of discomfort appear. Future work should compare such hybrid controllers against pure preference-driven optimization to test whether online gating can retain the attentional benefits reported by some participants while reducing the fatigue, uncertainty, and monitoring costs reported by others.

\subsection{Limitation and Future Work}
This study has several limitations that provide a foundation for future exploration of adaptive blur interfaces in real-world driving.

First, our implementation focused on object-based blur segmentation, applying blur selectively to predefined scene objects such as vehicles, buildings, and advertisements.
We did not explore screen-space blur regions, spatially continuous areas of blur independent of object boundaries.
Future studies could investigate blur in screen space to better understand how the spatial extent and gradient of blur influence drivers' performance, thereby providing a comprehensive design space for blur techniques.

Second, preferences for blur, baseline, or defined blur were collected after participants had experienced the corresponding conditions. These judgments are therefore retrospective evaluations rather than independent grouping variables. Consequently, preference-based patterns should be interpreted as descriptive evidence about how participants made sense of the interaction, not as causal evidence that prior preference determined whether blur improved or impaired driving performance.

Third, recent accident analyses and media reports suggest that many distraction-related crashes result from sudden or unexpected events occurring while the driver’s attention is diverted~\cite{Drivermonitorreview, WHOReport}.
Our current experimental design did not include such event-driven situations.
Future research could incorporate sudden hazards or time-critical events to examine how visual blur affects drivers’ reactions under realistic time pressure.

Fourth, this study was conducted in a controlled VR simulation environment.
Although VR enables safe and repeatable experimentation, real-world validation—particularly through on-road AR-HUD prototypes—will be essential to confirm the ecological validity of blur-based visual simplification techniques.

Fifth, our study concentrated exclusively on the blur-based visual simplification technique. We did not compare it against other established visual simplification methods, such as transparency modulation, saturation reduction, or DR. Comparative studies across these techniques would help clarify the relative strengths, perceptual impacts, and situational suitability of different approaches to visual load reduction.

Finally, our study did not incorporate eye-tracking measurements or direct assessments of hazard reaction time. As a result, we were unable to observe detailed changes in visual attention (e.g., gaze allocation, fixation patterns) or immediate behavioral responses to critical events. These measures are particularly important when evaluating visual guidance techniques, as blur may influence where drivers look or how quickly they respond, even when aggregate driving performance metrics remain unchanged. Future work should integrate eye-tracking and event-based reaction time analysis to provide a more sensitive and mechanistic understanding of how blur affects visual attention and safety-critical behavior.

\section{Conclusion}
This work examined whether personalized visual blur can proactively mitigate distracted driving by simplifying the visual environment. Through two VR user studies and a HITL-MOBO-based personalization framework, we found that visual blur does not improve overall driving performance. Instead, drivers exhibited pronounced individual differences in how they perceived and responded to blur. Pareto-front analyses revealed that optimal blur configurations varied widely across participants, indicating that no single design satisfies all drivers. Even with personalized Pareto-optimal solutions, personalized blur did not translate into measurable gains in driving safety. Qualitative feedback helps explain this discrepancy: while blur reduced visual clutter for some drivers, it increased perceptual uncertainty, mental effort, or fatigue for others, particularly in complex or distracted scenarios. These findings highlight fundamental limits of personalization and optimization in safety-critical tasks. While human-in-the-loop optimization effectively exposes individual preferences and trade-offs, it cannot eliminate perceptual conflicts inherent to visual blur. Future in-vehicle interfaces should therefore treat visual simplification as a conditional, adaptive intervention rather than a universal solution, dynamically adjusting or disabling it based on driving context and driver state.

\section*{Open Science}
The Unity application will be made available upon request, and the collected data is available via \url{https://anonymous.4open.science/r/BlurDriving_Data-5B00/}.

\begin{acks}
This work was supported by JST ASPIRE, Grant Number JPMJAP2401, and JST CRONOS, Grant Number JPMJCS24K8, Japan. Additional support was provided by a Canon Research Fellowship.
\end{acks}

\bibliographystyle{ACM-Reference-Format}
\bibliography{software, sample-base}

\appendix

\begin{figure}[ht!]
    \centering
    \small
        \centering
        \includegraphics[width=\textwidth]{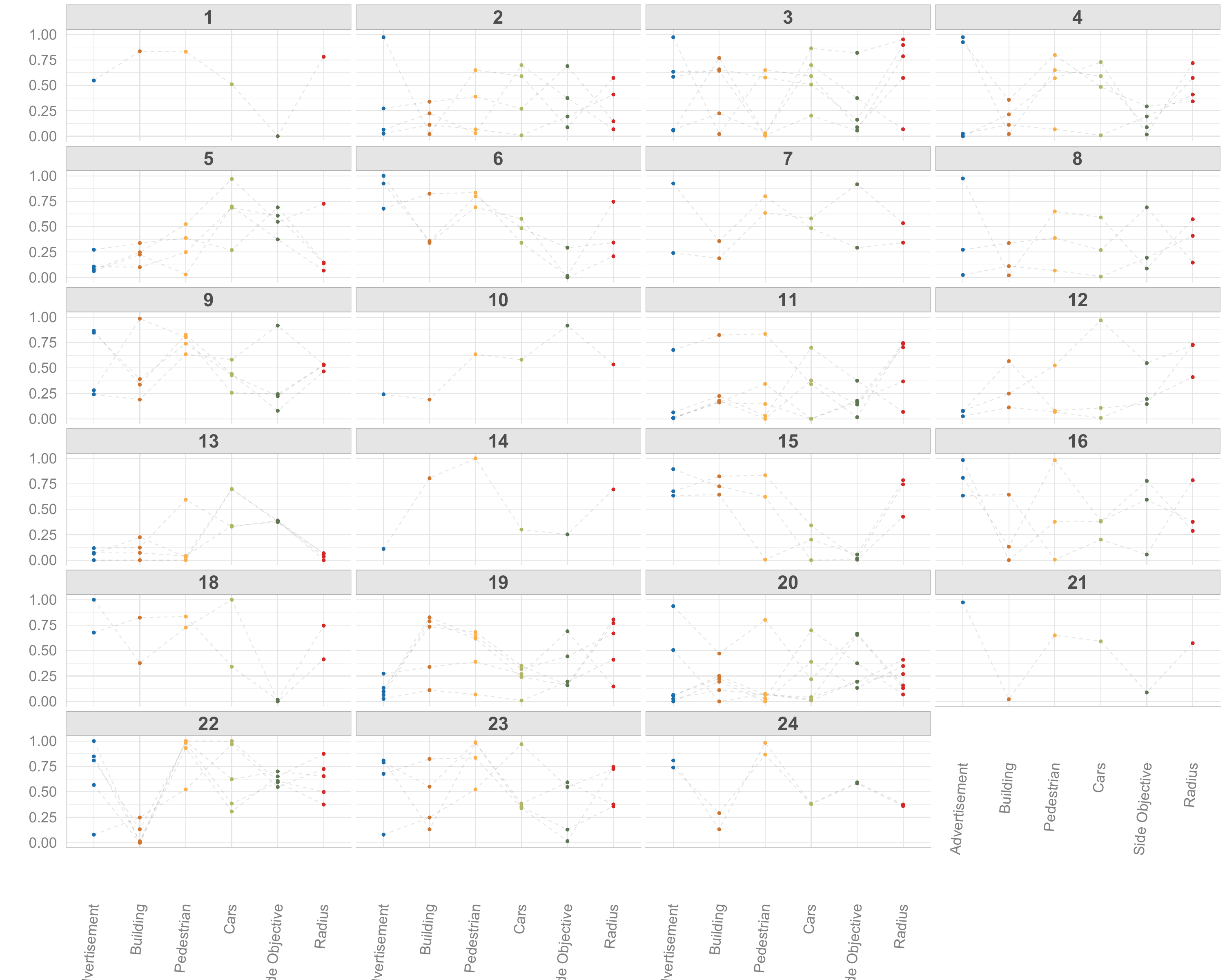}
    \caption{User Study 1: Pareto true values for all parameters per participant.}
    \label{fig:values_all_Per_person}
    \Description{}
\end{figure}

\begin{figure}[ht!]
    \centering
    \small
        \centering
        \includegraphics[width=\textwidth]{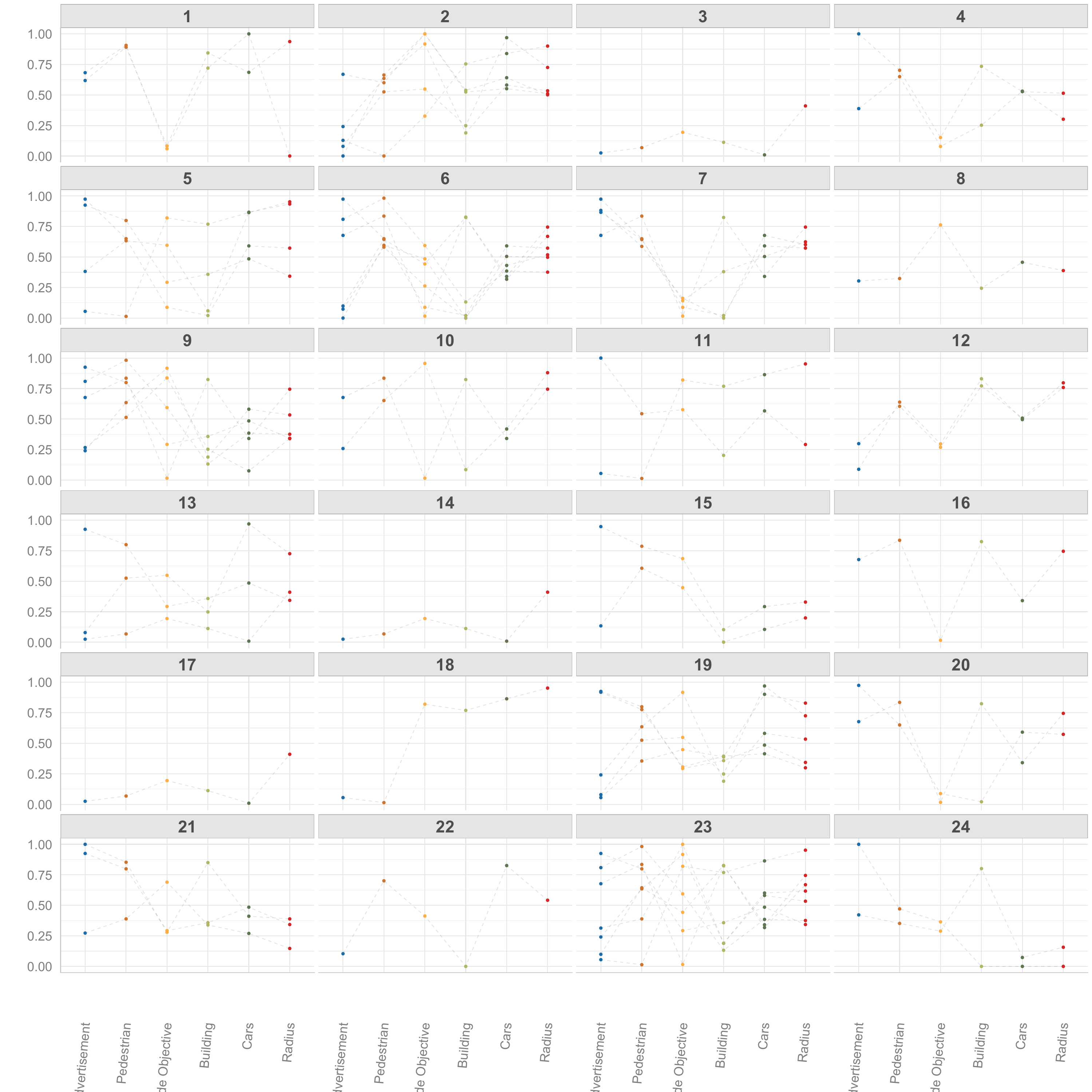}
    \caption{User Study 2: Pareto true values for all parameters per participant.}
    \label{fig:study2_values_all_Per_person}
    \Description{}
\end{figure}

\end{document}